\documentclass[journal]{IEEEtran}

\usepackage{color}
\usepackage{cite}
\usepackage[tight,footnotesize]{subfigure}
\usepackage{array}
\usepackage{tabu}
\usepackage{tabularx}
\usepackage{longtable}
\usepackage[pdftex]{graphicx}
\usepackage{amsmath}
\usepackage{pifont}
\usepackage{gensymb}
\usepackage{booktabs}
\usepackage{bm}
\usepackage[tight,footnotesize]{subfigure}
\usepackage[tracking=true]{microtype}
\usepackage{multirow}
\usepackage{algorithm}
\usepackage{algorithmic}
\usepackage{float}
\usepackage{textcomp}
\usepackage{gensymb}

\begin{document}
\title{A Low Power In-Memory Multiplication and Accumulation Array with Modified Radix-4 Input and Canonical Signed Digit Weights}
\author{Rui Xiao, ~\IEEEmembership{Student Member,~IEEE}, Kejie Huang, ~\IEEEmembership{Senior Member,~IEEE}, Yewei Zhang, ~\IEEEmembership{Student Member, ~IEEE} and Haibin Shen%
\thanks{The authors are with the College of Information Science \& Electronic Engineering, Zhejiang University, 38 Zheda Road, Hangzhou, China, 310027, email: xiaor@zju.edu.cn; huangkejie@zju.edu.cn; yeweizhang@zju.edu.cn; shen\_hb@zju.edu.cn.}
\thanks{K. Huang and H. Shen are also with Zhejiang Lab, Building 10, China Artificial Intelligence Town, 1818 Wenyi West Road, Hangzhou City, Zhejiang Province, China.}}

\markboth{IEEE TRANSACTIONS ON CIRCUITS AND SYSTEMS-I}%
{Shell \MakeLowercase{\textit{et al.}}: Bare Demo of IEEEtran.cls for IEEE Journals}
\maketitle

\begin{abstract}
A mass of data transfer between the processing and storage units has been the leading bottleneck in modern Von-Neuman computing systems, especially when  used for Artificial Intelligence (AI) tasks. Computing-in-Memory (CIM) has shown great potential to reduce both latency and power consumption. However, the conventional analog CIM schemes are suffering from reliability issues, which may significantly degenerate the accuracy of the computation. Recently, CIM schemes with digitized input data and weights have been proposed for high reliable computing. However, the properties of the digital memory and input data are not fully utilized. This paper presents a novel low power CIM scheme to further reduce the power consumption by using a Modified Radix-4 (M-RD4) booth algorithm at the input and a Modified Canonical Signed Digit (M-CSD) for the network weights. The simulation results show that M-Rd4 and M-CSD reduce the ratio of $1\times1$ by 78.5\% on LeNet and 80.2\% on AlexNet, and improve the computing efficiency by 41.6\% in average. The computing-power rate at the fixed-point 8-bit is 60.68 TOPS/s/W.
\end{abstract}


%
\IEEEpeerreviewmaketitle

\section{Introduction}
\IEEEPARstart{A}{long} with computer technology unceasing development, the Artificial Intelligence (AI) has been widely applied in various fields to perform specific tasks, such as transportation, education, healthcare, security, finance, etc \cite{LeCun2015}. With the support of massive data and high-performance hardware, the Deep Neural Network (DNN), a particular kind of machine learning, achieves excellent power and flexibility by learning to represent the world as a nested hierarchy of concepts \cite{liu2017survey}. However, due to the limited on-chip memory and memory bandwidth, a mass of intermediate data generated by DNN has to be transferred frequently between the separated computing units and storage units in conventional von-Neumann machines, resulting in a  tremendous amount of power and propagation delay,  which is treated as the ``Von Neumann bottleneck''\cite{burr2017neuromorphic}.

Inspired by the cranial nerve structure and information processing mechanism from the brain science research, the artificial intelligence and system are breaking through the conventional computing architecture and promoting to the next generation of computing paradigm. The computing-in memory \cite{xia2019memristive} scheme is formed by a large number of interconnected low-power computing units (neurons) and re-configurable storage units (synapses), which can perform the  Multiplication-and-Accumulation (MAC) operations in the memory to significantly reduce the data movement. The emerging Resistive Random Access Memory (RRAM) is one of the best candidates in CIM design \cite{huang2014,prime,pan2018mlc}. The resistance value of the memory can be used for weight storage and MAC operation. The memory cells are organized into crossbar arrays for high density storage, low power consumption, and fully parallel computing \cite{xia2019memristive}. 

Analog computing with multi-level resistive memory is widely used to achieve a massive parallel low power computing \cite{Irm2017multi,hu2018mem,Giacomin2018robust}. However, data storage and transmission between computing cores require digital signals since analog signals are sensitive to noise. Most architectures require Digital-to-Analog Converters (DACs) and Analog-to-Digital Converters (ADCs) at the interface, which consume large area and high power consumption. Moreover, most of them have overlooked the defects of the resistive Non-volatile Memory (NVM), such as nonlinearity, stochasticity, asymmetry, etc \cite{chen2011variability}. To address the issues mentioned earlier, \cite{zhang2020robust} proposes to use multiple binary RRAMs to emulate one synapse. Moreover, DACs are also moved to neurons to reduce the high driving power and the non-linearity caused by the analog input voltage. However, high-performance amplifiers are used  to achieve high computing speed and 8-bit resolution, resulting high power consumption. The high power dissipation amplifiers are removed in \cite{zhang2020regulator} by regulating the voltage before the passive integral neurons. However, 2's complementary code is used in synapses, resulting in balanced `1's and `0's. Moreover, the uncertainty of the memory resistance in the MAC array with 2's complementary code may cause a big jump between the most negative and the most positive values. Differential weights with Modified Canonical Signed Digital (M-CSD) are proposed to leverage the unbalanced `1's and `0's in weights to address above issues. Modified radix-4 (M-RD4) booth algorithm is also used to further reduce the percentage of `1's in the computation. The simulation results show that the total power consumption is reduced by more than 41.55\%. The performance-power ratio is 57.53 TOPS/s/W with 8b precision. The main contributions of this paper include:



\begin{enumerate}[\IEEEsetlabelwidth{3)}]
  \item The inputs are encoded with M-RD4 codes, the amount of `1's is halved since the encoding length of radix-4 booth codes is only half of binary encodes. 
  \item The weights are stored differentially with the M-CSD  code to significantly reduce the number of `1's, which can complete the MAC operation with 41.55\% less power computation.
  \item Differential charge redistribution passive integrator and Successive Approximation Register (SAR) ADC are proposed to enable in-memory computing with M-RD4 and M-CSD algorithms. 
\end{enumerate}

The rest of the paper is organized as follows: Section II introduces the related works of the resistive non-volatile memory based in-memory computing circuits and architectures. Section III discusses the detailed design of the proposed CIM core, including M-RD4, M-CSD, the integration scheme to perform MAC operations, and the corresponding circuits. Section IV provides the circuit level and system level simulation results. Finally, the conclusion is drawn in Section V.

\section{Related Works}
Computing near memory and computing in memory are the two typical schemes to shorten the distance between the processing and storage units. Computing near memory such as IBM TrueNorth\cite{merolla2014million} and Intel Loihi\cite{davies2018loihi} can only access the memory by one row each time, thus the processing speed is minimal. Besides, 
excessive charge and discharge of the bit lines will cost high power consumption. The CIM scheme could simultaneously access the whole array to perform the MAC operations, 
thus significantly reducing the latency and power for computing and memory access. Resistive NVMs such as memristor\cite{zhang2018neuromorphic}, Phase Change Memory (PCM)
 \cite{huang2014,burr2015experimental}, and RRAM \cite{Giacomin2018robust} are the potential candidates to achieve the high-density CIM schemes. Since all resistive NVMs have high 
 write power, the  network weights are usually trained offline on the server and then sent to the CIM cores for inference. The CIM schemes can be divided into two groups: CIM with 
 analog memory and input signals, and CIM with digitized memory and input signals.  

\subsection{Analog Computing-in-Memory}
In the analog CIM schemes, the multiplication is usually achieved by multiplying the conductance of the multi-level memory and input analog voltage based on  Ohm's law
\cite{sheridan2017sparse,Jiang2018}, which will output the current. The accumulation is usually done by converging output currents from different multiplications based on  Kirchhoff’s 
Current Law (KCL). The analog signals are difficult to be preserved and also sensitive to noises. Therefore, the converged current has to be converted to voltage signals for 
analog-to-digital conversion. The digital inputs are also converted to analog signals for analog computing. The DACs and ADCs will consume enormous power and area, significantly
limiting the efficiency of the scheme.

\begin{figure}
  \centering
  \includegraphics[width=0.45\textwidth]{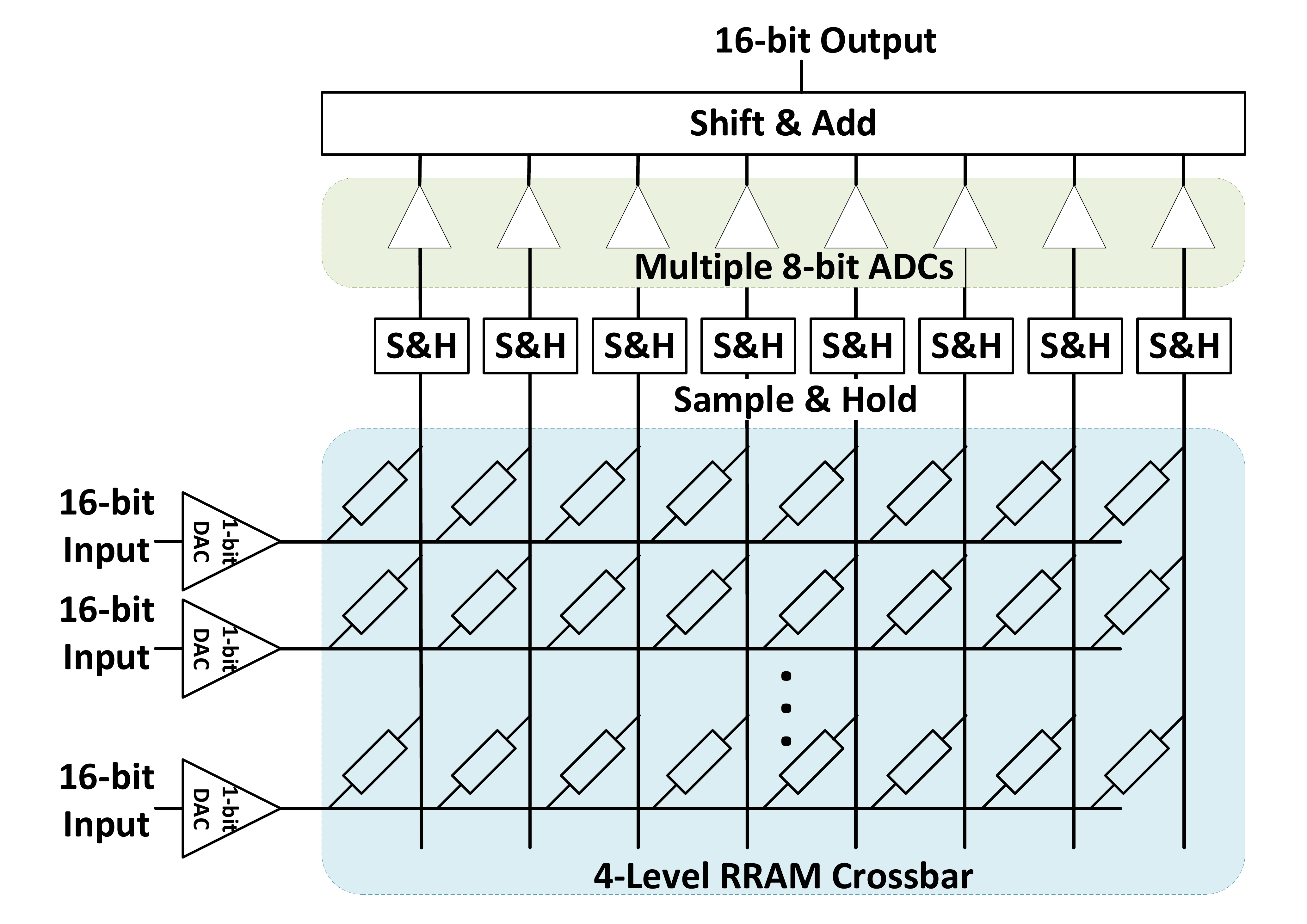}
  \caption{The CIM core of the ISAAC architecture. The S/H voltage of each column is quantized by an individual 8-bit ADC at each cycle, and then the quantized results are 
  shifted and added to achieve 16-bit output for MAC operations.}
  \label{Fig:ISAAC}
\end{figure}

A. Shafiee \textit{et al.} \cite{ISAAC} proposed the RRAM-based ISAAC scheme to perform 16-bit fixed-point MAC operation for CNN inference, where eight 4-level RRAM cells are used to 
store one 16-bit weight. As shown in Fig. \ref{Fig:ISAAC}, it takes 16 cycles to perform the 16-bit digital-to-analog conversion by 1-bit DACs instead of 16-bit high-cost DACs in 1 
cycle. In each cycle, the analog outputs are converted to digital signals by eight 8-bit ADCs, which are then shifted and added to generate 16-bit output. X. Qiao \textit{et al.} 
\cite{Atomlayer} proposed AtomLayer to support 16-bit fixed-point CNN training and inference. The AtomLayer accesses the ability of training by processing one network layer each time. B
esides, the data are redused to improve the efficiency. However, there are still some shortcomings in ISAAC and AtomLayer.

\begin{enumerate}[\IEEEsetlabelwidth{4)}]
  \item The S/H structure without an amplifier will seriously affect the analog computation accuracy due to the varying hold voltage.
    \item The accuracy after ADC is far less than 8-bit due to the nonlinearity of multi-level RRAM cells and the loss of precision.
  \item The shift-and-add operation will further reduce the accuracy because ADC's quantization error is magnified after the shift operation. 
  \item The eight ADCs and 16 cycles' conversion for 16-bit MAC operation leads to high power consumption.
\end{enumerate}

\begin{figure}
  \centering
  \includegraphics[width=0.45\textwidth]{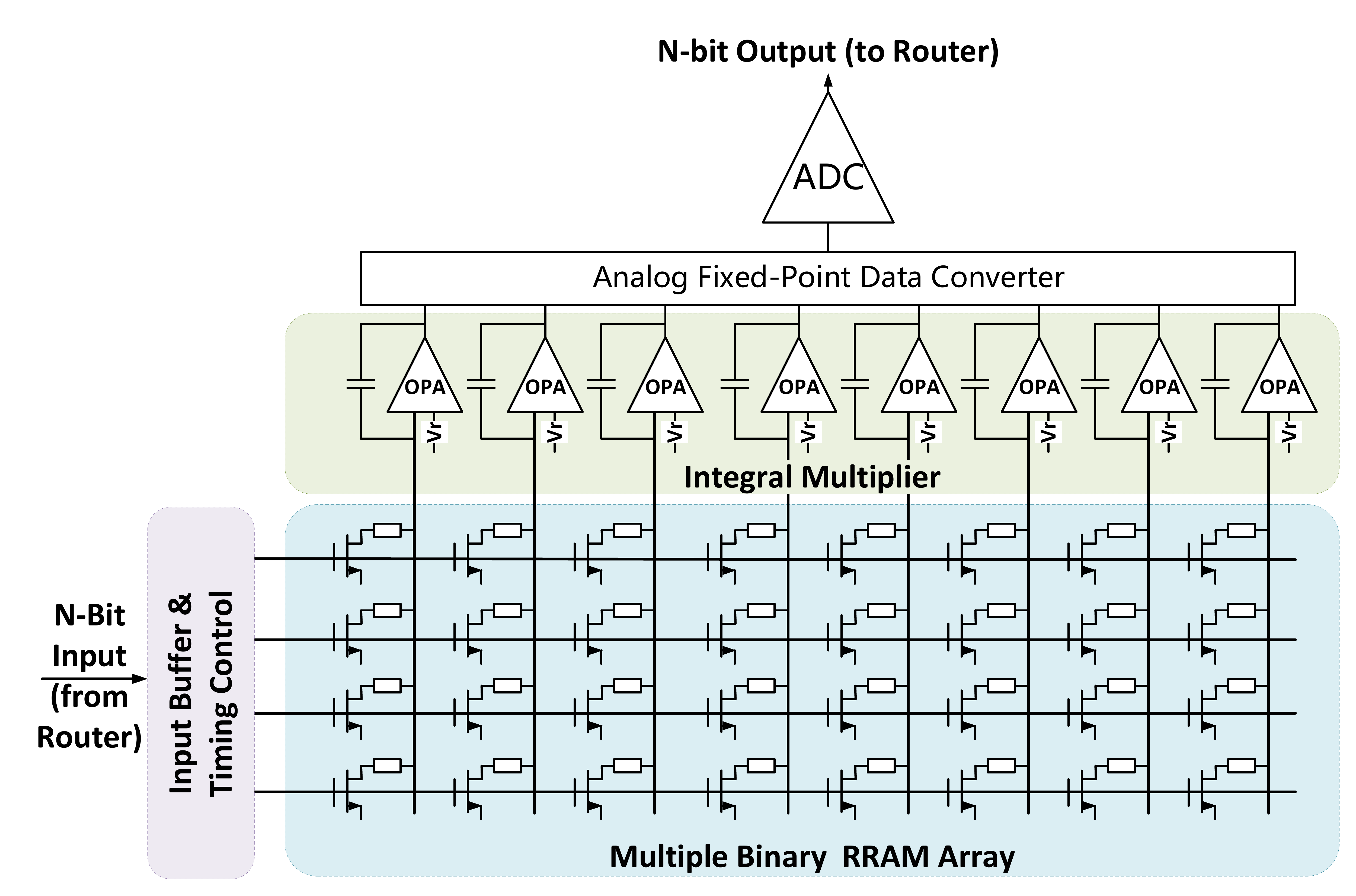}
  \caption{The CIM core of the MBRAI architecture. The n-bit input data are sequentially computed in the integral multiplier and weighted at the output neurons}
  \label{FIG:MBRAI}
\end{figure}

\subsection{Digitized RRAM based CIM Cores}
Several single level RRAM based CIM cores have been proposed to avoid the nonlinearity issue of multi-level RRAM. M. Courbariaux  \textit{et al.}\cite{courbariaux2016binarized} and 
M. Rastefari \textit{et al.}\cite{rastegari2016xnor} used binary weight and 1-bit input for the recognition tasks on MNIST and CIFAR-10 datasets. However, 1-bit weight and 1-bit input 
will lose a lot of information when applied to large networks. C. Xue \textit{et al.}\cite{Xue2020154A2} proposed a BL-IN-OUT (BLIOMC) scheme with Scrambled 2's Complement Weight Maping 
(S2CWM), which exploits 4-bit inputs by 4-level read voltage and 4-bit weight represented by four single-level RRAM cells. The Dual-bit-Small-Offset Current-mode Sense Amplifier 
(DbSO-CSA) with two $I_{REF}$ works as 2-bit ADC. It achieves an efficiency with 28.9 TOPS/s/W at 4-bit input and 4-bit weight. However, the structure of this design limits its 
application to some extent:
\begin{enumerate}[\IEEEsetlabelwidth{4)}]
  \item The 4-level read voltage $V_{RD}, 2/3V_{RD}, 1/3 V_{RD}, 0$ at the input will vary memory resistance during the read operation, which will affect the accuracy of MAC operation.
  \item The 2b sensing amplifier will greatly limit the total precision of the MAC output. Adding multiple outputs will average the quantization error and noise, but the increased 
  precision is halved.
  \item It needs multiple cycles to finish the Vector-Matrix operation, which will significantly reduce the computation speed.
\end{enumerate}

To address the issues mentioned above and further improve the energy efficiency, S. Zhang \textit{et al.}\cite{zhang2020robust} proposed a Multiple Binary RRAM with Active Integrator 
(MBRAI) core architecture. As shown in Fig. \ref{FIG:MBRAI}, multiple binary-RRAM cells are used to represent an 8-bit weight instead of a multi-level RRAM cell. The core uses binary 
code at the input instead of a time signal or analog signal. The n-bit data are sequentially computed in the integral multiplier and weighted at the output neurons. However, the 
amplifiers in the neurons are power-hungry components to achieve a wide dynamic range, which consume more than 95\% power in the scheme. The computing efficiency of the CIM core is 
limited to 0.61 TOPS/s/W. To address this issue, Y Zhang \textit{et al.} \cite{zhang2020regulator} proposed an 8-bit In Resistive Memory Computing Core with Regulated Passive Neuron 
and Bit Line Weight Mapping (RPN \& BLM) scheme.  RPN \& BLM uses passive integral circuits without amplifiers to decrease  power consumption. The regulators in the bit lines are used 
to improve the linearity of the integration process.

\begin{figure}
  \centering
  \includegraphics[width=0.45\textwidth]{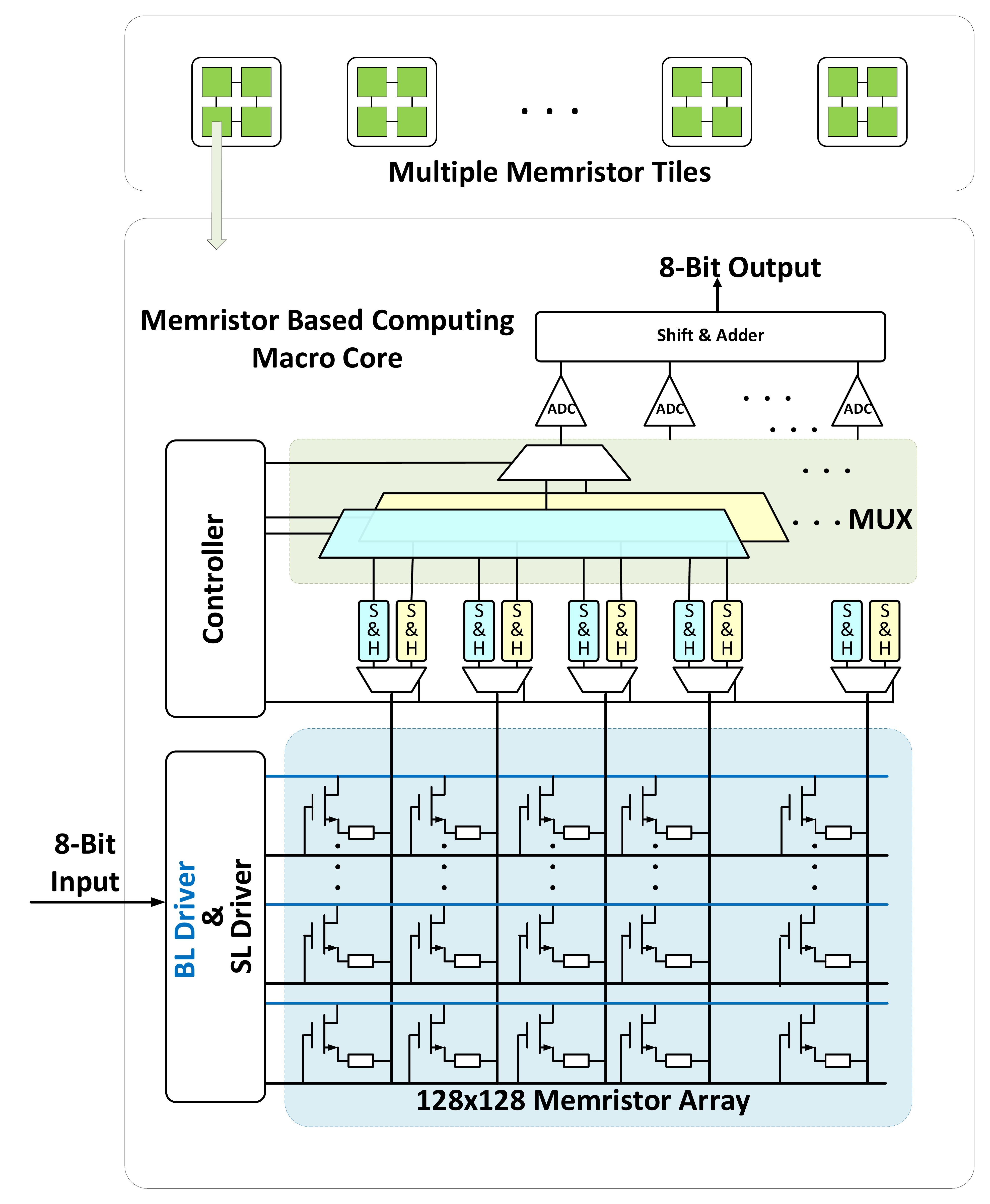}
 \caption{The architecture of the memristor-based neural processing unit and relevant circuit modules.}
  \label{FIG:nature}
\end{figure}

\subsection{Differential Weight based CIM Cores}
The uncertainty of the memory resistance in the MAC array with 2's complementary code may cause a big jump between the most negative value (i.e., 8'b10000000) and the most positive value (i.e., 8'b01111111). Differential weights \cite{prime,nair2017differential,Joshi2020} could be used to avoid this issue. Recently, P. Yao \textit{et al.} \cite{yao2020fully} proposed a memristor-based hardware system with reliable multi-level conductance states for a five-layer mCNN for MNIST digit image recognition (MBHS-mCNN). As shown in Fig. \ref{FIG:nature}, the neural processing unit consists of multiple memristor tiles and each tile contains four memristor cores. The MUX controller is used to select the positive and negative computing results. However, there are still some weaknesses in this scheme:
\begin{enumerate}[\IEEEsetlabelwidth{4)}]
  \item It requires 32 times of analog-to-digital conversions and Shift \& add operations to finish one MAC operation, which consume about 92.14\% energy in the system.  
  \item The quantization error is amplified by the shift and add operation, and thus it cannot achieve the desired precision.
\end{enumerate}


\section{Proposed In-memory Computing Core}
\begin{figure}
  \centering
  \includegraphics[width=0.45\textwidth]{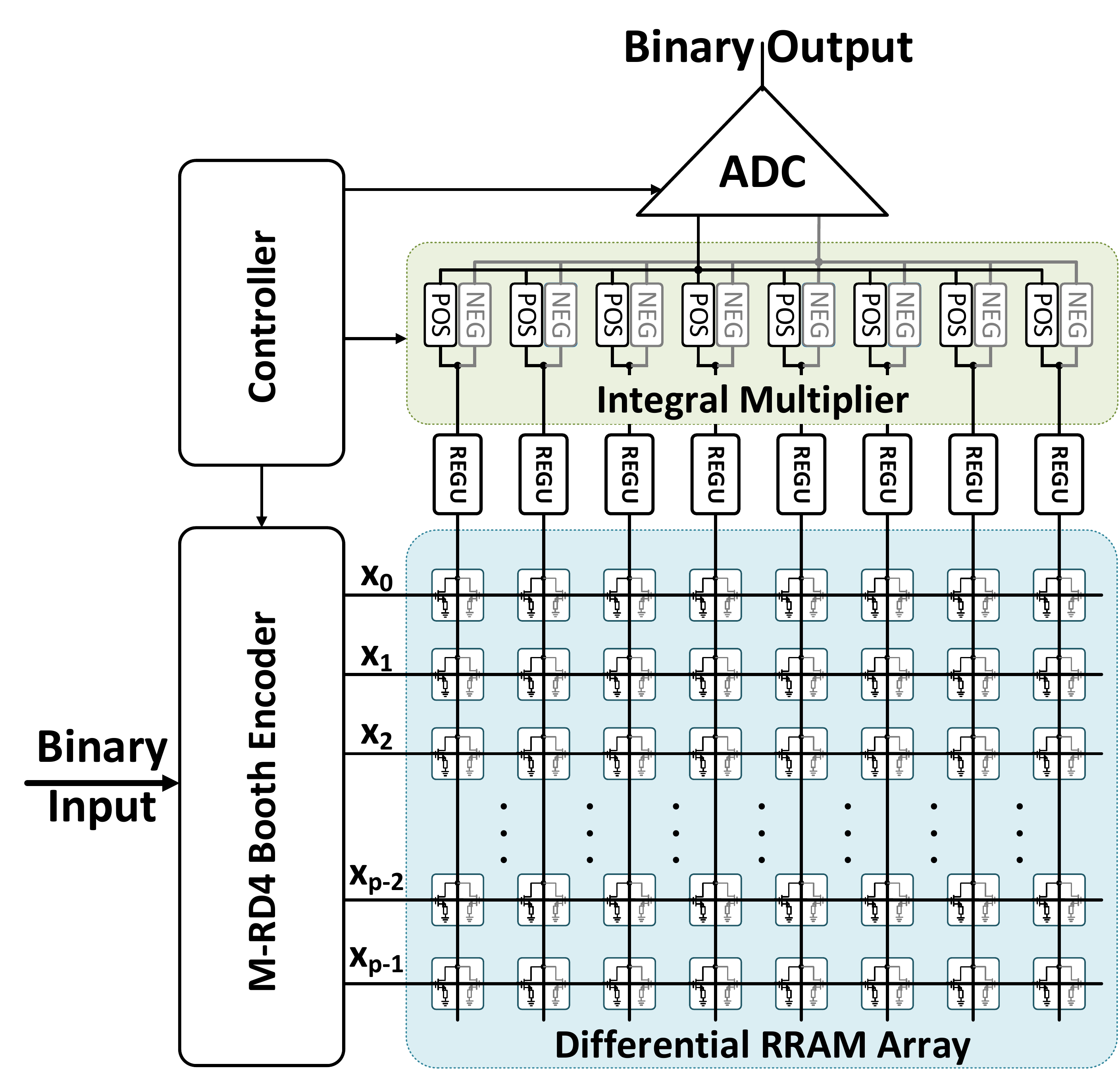}
  \caption{The overall architecture of OUR proposed CIM core.}
  \label{FIG:arch}
\end{figure}
In this paper, we propose a booth encoded differential core for low-power parallel MAC operations. M-RD4 and M-CSD algorithms are proposed at the input and weights respectively to 
reduce the power consumption of MAC operations. The overall structure of the proposed scheme is shown in Fig. \ref{FIG:arch}, which consists of six components, including M-RD4 generator,
differential RRAM array, regulator, integrator, controller, and differential ADC. To be simplified, only 8$\times$8 crossbar memory cells are illustrated in Fig. \ref{FIG:arch}. 
It can be extended to 8$\times$N$\times$N memory cells for the real application. Each memory cell is comprised of a 1R1T pair. The binary inputs are firstly converted to the stimulus 
of the CIM core by using an M-RD4 booth algorithm. The stimulus will turn on the transistor in 1R1T to generate the current to pass through RRAM cells and accumulated at the integrators 
to enable the massive parallel MAC computation. Regulator\cite{zhang2020regulator} is used before the integrator to minimize the voltage variation caused by the channel length 
modulation during the integration. Finally, the analog voltage at the neuron is converted to the digital signals using the charge redistribution differential SAR ADC. Only one 8-bit ADC 
is required by eight integrators for high density and low power. The details of each block will be introduced in the rest of this section.

\subsection{Modified Radix-4 Booth Code}
Unsigned fixed point data can be used as the CIM core input because there is no negative data after the ReLU activation function. The input data can be expressed as an n-bit unsigned 
fixed-pointed data $X_k$

 \begin{equation}\label{Eq:x}
   X_k = 2^{n-1}x_{k,n-1}+...+2^{i}x_{k,i}+...+2^{0}x_{k,0}
 \end{equation}

Radix-4 booth code \cite{booth1951signed} is a modified booth code used for high-speed and low-power computing, widely used to design the multipliers to halve the number of partial 
products. The algorithm of recoding an n-bit binary number (X) to a radix-4 booth number (Z) is as follows. Firstly append a ‘0’ to the right of the Least Significant Bit (LSB) of the 
X, and then extend the sign bit one position if necessary to ensure that n is even. After that, every three binary bits (with 1 bit overlap) are encoded as one radix-4 bit from the 
LSB to the Most Significant Bit (MSB).The eight cases of the radix-4 code are tabulated in Table \ref{Table:rd4}. By using the radix-4 algorithm, the length of the input code is 
halved (i.e. 01111111 is encoded to 200$\bar1$). The number of `1's can also be reduced compared with binary codes, which means the power consumption can be reduced since more 
multiplications can be bypassed in the MAC calculations.

\begin{table}
  \centering
  \caption{The Truth Table of Radix-4 Code and Proposed M-RD4 Code}
  \resizebox{0.45\textwidth}{22mm}{
  \begin{tabular}{|c|c|c|c|c|c|}
  \hline
  \multicolumn{4}{|c|}{Binary Bits} & Radix-4 Bit & M-RD4 Bit \\
  \hline
  $t_{i+3}$&$t_{i+2}$ & $t_{i+1}$ & $t_{i}$ &  & $z_j$ \\
  \hline
  0/1&0&0&0 &0 &0 \\
  \hline
  0/1&0&0&1 &1 &1 \\
  \hline
  0/1&0&1&0 &1 &1 \\
  \hline
  0&\multirow{2}{*}{0}&\multirow{2}{*}{1}&\multirow{2}{*}{1}  &\multirow{2}{*}{2} &2\\
  \cline{1-1}
  \cline{6-6}
  1& & & & & -2\\
  \hline
 0&\multirow{2}{*}{1}&\multirow{2}{*}{0}&\multirow{2}{*}{0}  &\multirow{2}{*}{-2} &2\\
  \cline{1-1}
  \cline{6-6}
  1& & & & & -2\\
  \hline
  0/1&1&0&1  &-1 &-1 \\
  \hline
  0/1&1&1&0 &-1 &-1 \\
  \hline
  0/1&1&1&1 &0 &0 \\
  \hline
  \end{tabular}
  }
  \label{Table:rd4}
\end{table}

 \begin{figure}
	\centering
	\subfigure[Radix-4]{
		\begin{minipage}[b]{0.22\textwidth}
			\includegraphics[width=1\textwidth]{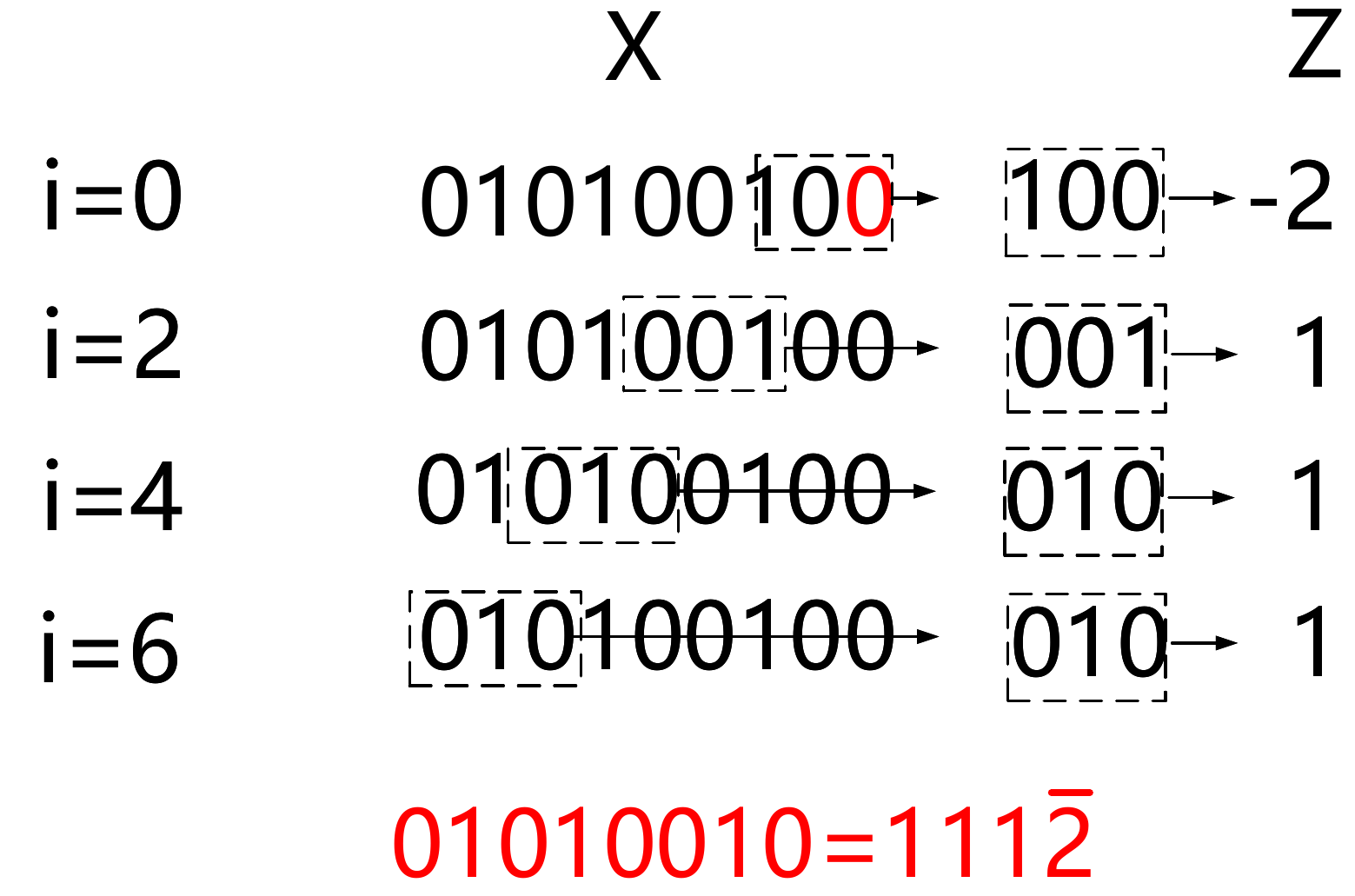}
    \end{minipage}
   
		\label{FIG:RD4}
  }
  \subfigure[M-RD4]{
    \centering
  \begin{minipage}[b]{0.22\textwidth}
   	\includegraphics[width=1\textwidth]{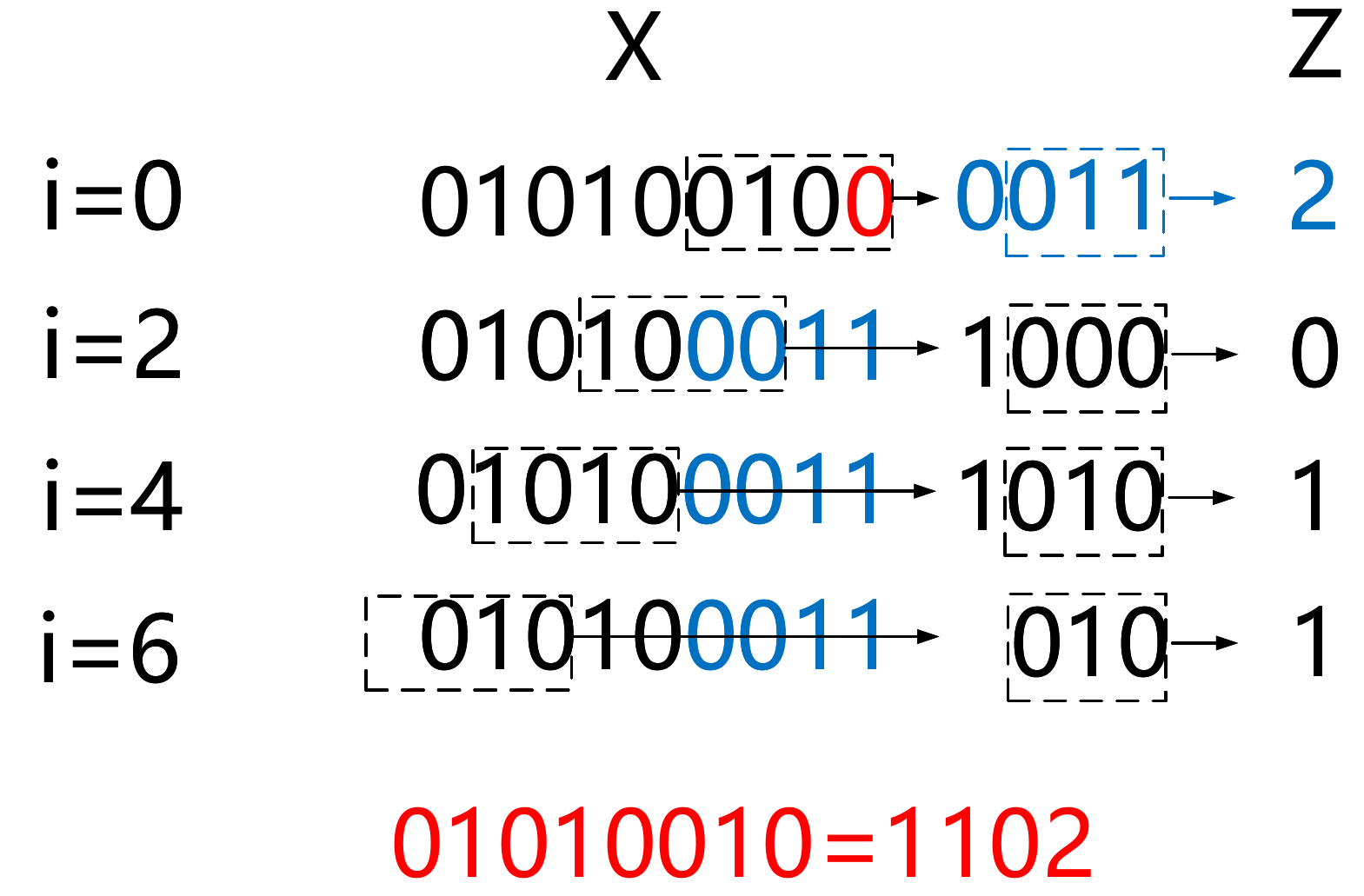}
    \end{minipage}
    \label{FIG:M_RD4}
    
  }
  
	\caption{The example of (a) radix-4 code, (b) M-RD4 code. The proposed M-RD4 code can effectively reduce the number of `1's in the input data to a minimum}
	\label{FIG:INPUT_CODE}
\end{figure}

%

However, the radix-4 code sometimes leads to more `1's than that in binary codes. Fig.\ref{FIG:RD4} shows an example to encode a binary code `01010010' to the radix-4 code `111$\bar2$',
where the number of `1's is increased in radix-4 code. To reduce the number of `1's in radix-4 code, we propose an M-RD4 code to get the least `1's at the input. The M-RD4 algorithm is
illustrated in Algorithm \ref{Alg:m_rd4} . The proposed M-RD4 algorithm will observe one more bit at the left. If the sequence is `0100', it will be turned into `0011'. If the sequence
is `1011', it will be turne into `1100'. After that, the right three bits will be encoded by using Eq (\ref{eq:rd4}). 
\begin{equation}
    z_j=-2t_{i+2}+t_{i+1}+t_i
    \label{eq:rd4}
\end{equation}
where $i=0,2,4,...,j=\frac{i}{2}$, $t_i$ is the $i_{th}$ bit of T, T is defined in Algorithm \ref{Alg:m_rd4}. The cases are tabulated in Table \ref{Table:rd4}. The M-RD4 code can 
further reduce the number of `1's in input data. Fig. \ref{FIG:M_RD4} is used as an example to illustrate our M-RD4 algorithm. The M-RD4 code of `01010010' is changed to `1102' instead 
of `111$\overline{2}$'.

\begin{algorithm}
\caption{M-RD4 Booth Code}
\label{Alg:m_rd4}
\begin{algorithmic}[1]
\REQUIRE Binary n-bit $X=x_{n-1}x_{n-2}...x_i...x_0$.
\ENSURE M-RD4 Booth Encoded m-bit data $Z=z_{m-1}z_{m-2}...z_j..z_0$, where $m=\left \lceil \frac{n}{2} \right \rceil$.

\STATE $//$ Extend a `0' at the most left to ensure that n is even
\STATE $//$ Append a `0' to the right of the Least Significant Bit (LSB) 
\IF{n is even}
\STATE T[n:0] $\Leftarrow$  $x_{n-1}x_{n-2}...x_i...x_0\bm{0}$
\ELSE
\STATE T[n+1:0] $\Leftarrow$  $\bm{0}x_{n-1}x_{n-2}...x_i...x_0\bm{0}$
\ENDIF
\STATE i $\Leftarrow$ 0, j $\Leftarrow$ 0
\WHILE{$i \leq n-2$}
\STATE $//$ Observe one more bit per time and transfer the sequence if necessary.
\IF{$t_{i+3}t_{i+2}t_{i+1}t_0 == `0100'$}
\STATE $t_{i+3}t_{i+2}t_{i+1}t_0 \Leftarrow `0011'$
\ELSIF{$t_{i+3}t_{i+2}t_{i+1}t_0 == `1011'$}
\STATE $t_{i+3}t_{i+2}t_{i+1}t_0 \Leftarrow `1100'$
\ENDIF
\STATE $//$ Get the M-RD4 code Z from LSB to MSB.
\STATE  $z_j \Leftarrow -2t_{i+2}+t_{i+1}+t_i$
\STATE $i \Leftarrow i+2$
\STATE $j \Leftarrow j+1$
\ENDWHILE
\RETURN Z
\end{algorithmic}
\end{algorithm}



\begin{figure}
  \centering
  \includegraphics[width=0.5\textwidth]{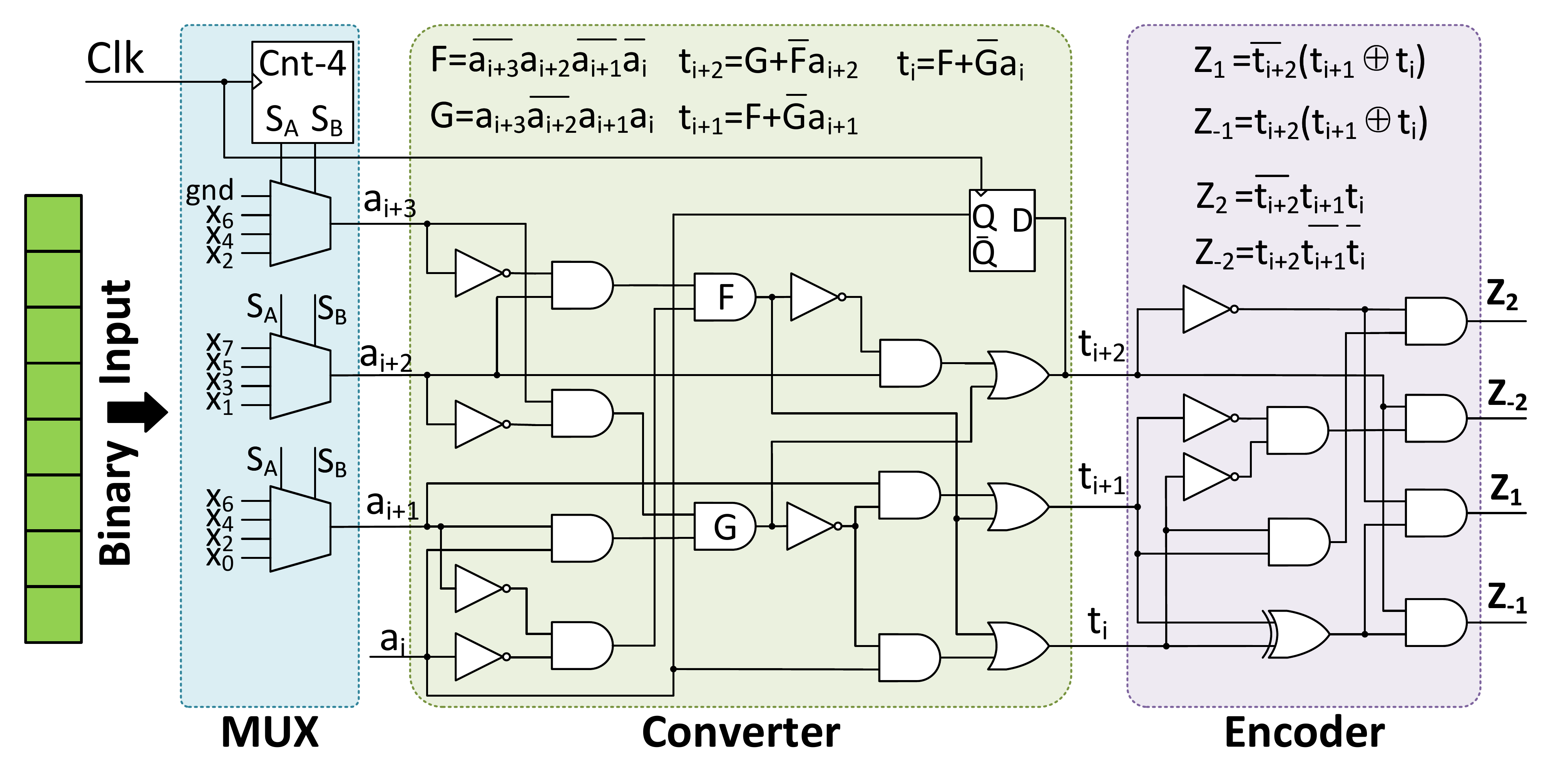}
  \caption{The circuit implementation of the proposed M-RD4 booth recoding circuit. The MUX block selects the bits of the binary input, and the converter processes the input under the 
  rules of the proposed M-RD4 algorithm. The encoder recodes the input to M-RD4 codes and outputs it into the neuron circuit.}
  \label{Fig:boothcir}
\end{figure}

Fig. \ref{Fig:boothcir} shows the M-RD4 booth recoding circuit implementation, which is composed of the MUX block, converter block and encoder block. The MUX block consists of three 
4-to-1 multiplexers and one quaternary counter. The counter generates the control signals ($S_A$ and $S_B$) to select the output of each multiplexer. In this way, the MUX block outputs 
the raw data for M-RD4 from the LSB to MSB. The converter block converts the raw data for encoding according to the M-RD4 algorithm. $a_{i+3}, a_{i+2}, a_{i+1}$ are the outputd of the 
MUX block, and $a_i$ is generated by the converter. As shown in Fig.\ref{FIG:M_RD4}, in the first clock, $a_i=0$, and then $a_i$ is determined by the output of the converter($t_{i+2}$) 
in the last clock. Therefore, $a_i$ is `0,0,0' in the next three clocks.  According to Algorithm \ref{Alg:m_rd4}, we set 
\begin{gather}
  F=\overline{a_{i+3}}a_{i+2}\overline{a_{i+1}}\bar{a_i} \\
  G=a_{i+3}\overline{a_{i+2}}a_{i+1}a_i 
  \label{eq:FG}
\end{gather}
then we can get the output of the converter block
\begin{gather}
  t_{i+2}=G+\overline{F}a_{i+2}\\
  t_{i+1}=F+\overline{G}a_{i+1}\\
  t_i=F+\overline{G}a_i
\end{gather}
The output of the converter is sent to the encoder block for recoding. The 3-bit binary codes are recoded to 1-bit M-RD4 code by combination circuit according to Table \ref{Table:rd4}.
The encoder output log can be shown as
\begin{gather}
  Z_2= \overline{t_{i+2}}t_{i+1}t_i \\
  Z_{-2}=t_{i+2}\overline{t_{i+1}} \bar{t_i}\\
  Z_1 = \overline{t_{i+2}}(t_{i+1} \oplus t_i) \\
  Z_{-1} = t_{i+2}(t_{i+1} \oplus t_i) 
 \end{gather}
where $Z_{2}$, $Z_{-2}$, $Z_1$, and $Z_{-1}$ represent four values of $z_j$ (2, -2, 1, -1) in Table \ref{Table:rd4}. When $z_j$ is encoded to zero, the multiplication result is always 
zero. Therefore, there are only four output terminals from the combination logic circuit, and only one of them will be activated at a time. If $z_j = 1$, then the voltage of $Z_1$ is 
high and the others are low, and the other cases can be speculated.

To make the M-RD4 code and its corresponding circuit clearer, we use the binary code `01010010' as an example. In the first clock, $S_A=0,S_B=1$, then $a_{i+3}a_{i+2}a_{i+1}=x_2x_1x_0(010)$, and $a_i=0$. According to Eq (\ref{eq:FG}), we can get $F=1,G=0$. The outputs ($t_{i+2}t_{i+1}t_i$) of the converter are 011. The M-RD4 result is 2, thus $Z_2=1$, $Z_{-2}=0$, $Z_{1}=0$, and $Z_{-1}=0$. In the second clock, $S_A=0,S_B=1$, then $a_{i+3}a_{i+2}a_{i+1}a_i=x_4x_3x_2Q(1000)$, where Q equals $t_{i+2}$ at the last clock. $F=0$, $G=0$, then $t_{i+2}t_{i+1}t_i=000$, therefore all of the outputs are 0 . In the third clock, $S_A=1$, $S_B=0$, then $a_{i+3}a_{i+2}a_{i+1}a_i=x_6x_5x_4Q(1010)$. $F=0$, $G=0$, then $t_{i+2}t_{i+1}t_i=010$, therefore $Z_1=1$. In the fourth clock, $S_A=1$, $S_B=1$, then $a_{i+3}=gnd(0)$, and $a_{i+2}a_{i+1}a_i=x_7x_6Q(010)$. $F=0$, $G=0$, then $t_{i+2}t_{i+1}t_i=010$. According to the third clock, $Z_1=1$. Therefore, the M-RD4 output is `1102'. The four output bits, which are either at VDD or ground, are directly used in the in-memory computing. The weights of 1, -1, 2, and -2 will be employed in the neuron circuit, which will be discussed in Section III. C.

\subsection{Modified CSD Weights}
\begin{figure}
  \centering
  \includegraphics[width=3.5in]{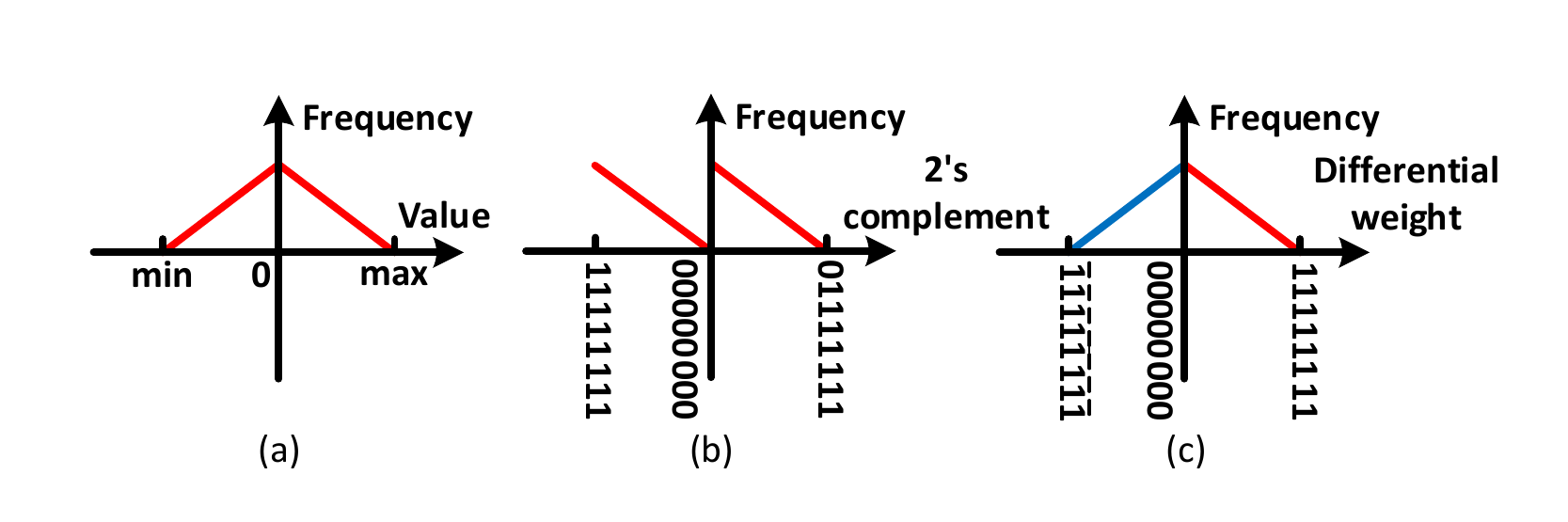}
  \caption{The simplified distribution curve of (a) weights, (b) weights in 2's complement, (c) differential weight.}
  \label{Fig:dist}
\end{figure}

2's complementary code representation is widely used in the arithmetic logic and operation. However, it may not be the best form to minimize the power consumption for the neural network
computing. Fig. \ref{Fig:dist}(a) shows the simplified distribution curve of the weights in a neural network. In unpruned DNN networks, the weight values often follow a normal 
distribution. Similarly, the inputs follow a half-normal distribution, because all negative values have been forced to be zero after the ReLU activation function. If the weights and 
inputs are qualified to 8-bit binary data, there are 40\% - 50 \% of `1's in the weights and about 20\% - 30\% of `1's in the inputs. If 2's complement is used, as shown in 
Fig. \ref{Fig:dist}(b), the  number of `1's and `0's will be balanced and the probability of $1\times1$ is about 10\%, which is not optimized for low power computing. What's more, 
the 2's complementary may cause a big jump between the most negative value (10000000) and the most positive value (01111111) due to the uncertainty of the memory resistance. The leap 
will significantly influence the accuracy of in-memory computing.

\begin{figure}
  \centering
  \includegraphics[width=2.5in]{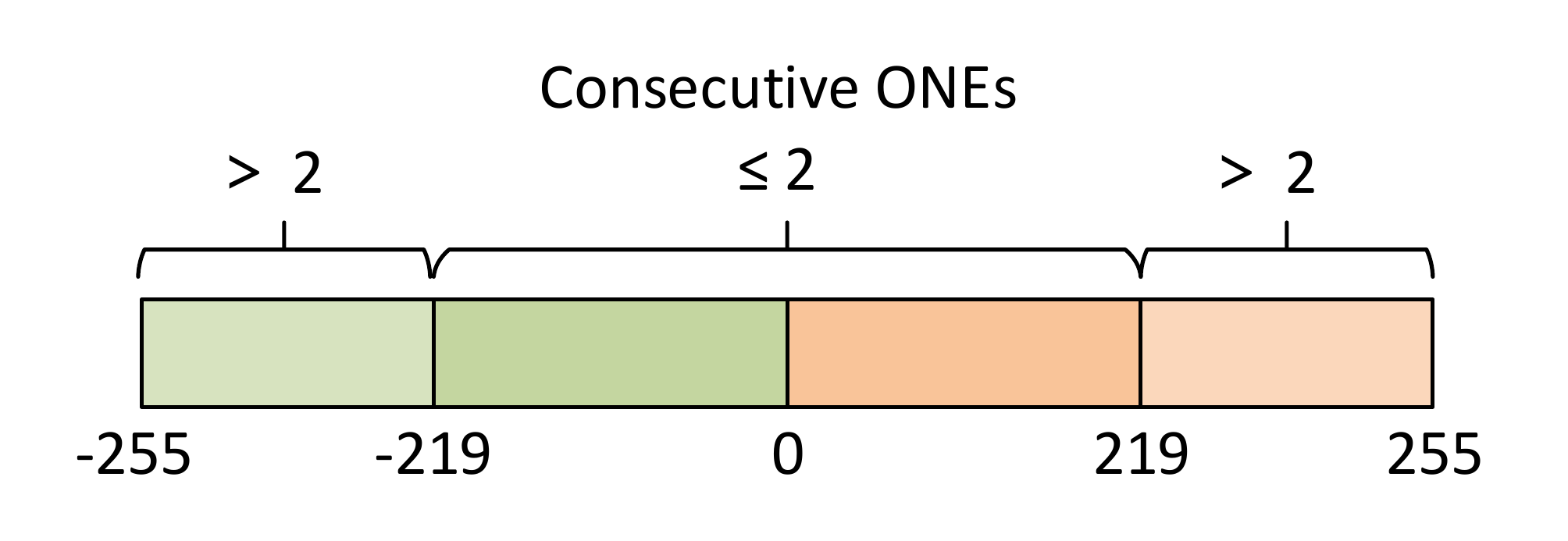}
  \caption{The data representation of our proposed differential weight system.}
  \label{Fig:csd}
\end{figure}

Differential weights can be used to address the above mentioned issues, which can be represented as
\begin{equation}
  \label{EQ:diff}
  w = w_{p} - w_{n} = 2^{n-1}(b_{n-1}-c_{n-1})+...+2^{0}(b_{0}-c_{0})
\end{equation}
where $w_p$ and $w_n$ are the unsigned number representation, and $b_{i}$ and $c_i$ are the bits in the positive part and negative part of a weight, respectively. For example, 
$w_p$ = 8'b00000000 and $w_n$=8'b01110111 represent weight -119. As shown in Fig.\ref{Fig:dist}(c), the red line indicates a positive value, and the blue line indicates a negative value.
The digits 1 and $\overline{1}$ are placed in the positive and negative parts of the weight, respectively. In this way, the majority of bits in the weights are 0, which could bypass 
the in memory computing to save the power consumption by around 50\%. 

However, it doesn't fully utilize both parts of a differential weight. If we could represent W with fewer non-zero digits, we could reduce the in-memory computing power consumption. 
CSD representation \cite{CSD} is widely used to reduce the non-zero digits by introducing a new digit $\overline{1}$ into the number to form a ternary number system. The pair $b_i$ and 
$c_i$ in Eq (\ref{EQ:diff}) can be used to represent the digit set \{1,0,$\overline{1}$\} for a CSD code. A simple approach to encode a binary code to a CSD code is to search the 
binary code from LSB to MSB, find a string of `1's followed by `0' (i.e. 0111), and replace them with the CSD representation (1000$\bar{1}$). The process may need to be repeated several
 times to make sure there is no string of `1's. CSD representation still suffers from some shortcomings:
\begin{enumerate}[\IEEEsetlabelwidth{4)}]
  \item In a CSD number, two consecutive non-zero bits are not allowed. Thus the maximum value of 8-bit CSD is limited to 170 (10101010). For those 8-bit binary numbers greater than 170,
  an extra bit is needed to represent them in CSD representation.
  \item For string `011', CSD representation (10$\bar{1}$) doesn't reduce the number of `1's.
\end{enumerate}

An M-CSD representation is proposed to address the above issues. The strings `11' and `$\Bar{1}\Bar{1}$' are allowed in M-CSD. The main idea of M-CSD is shown in Algorithm \ref{Alg:m_CSD}.
Strings containing three or more `1's will be replaced by 10...0$\overline{1}$ and  three or more `$\overline{1}$'s will be replaced by $\overline{1}0...01$. If the MSB of the binary 
code is contained in a string, then the string will not be replaced with the M-CSD representation. In this way, the maximum value is extended to 219 (11011011). As shown in 
Fig. \ref{Fig:csd}, to achieve the same range as the binary code, more consecutive `1's will be allowed if the weight is greater than 219 or smaller than -219.  In this way, the M-CSD 
code perfectly fits the differential weight scheme. To comply with the CSD design rule, $w_p$ and $w_n$ in the above example will be changed to 8'b00001001 and 8'b10000000, respectively.
Therefore, the number of `1's is significantly reduced.

\begin{algorithm}
\caption{Modified CSD Representation}
\label{Alg:m_CSD}
\begin{algorithmic}[1]
\REQUIRE n-bit differential $W_i=w_{n-1}w_{n-2}...w_0$.
\ENSURE n-bit modified CSD $W_i=w_{n-1}w_{n-2}...w_0$.
\STATE Flag $\Leftarrow$ 0 $//$ Mark the string containing the MSB.
\STATE  $i \Leftarrow n-1$   
\STATE $j \Leftarrow 0$ 
\STATE   $k \Leftarrow 0$ 
\STATE $//$String containing MSB will not be replaced.
\WHILE{i \textgreater 0 \& Flag == 0}
\IF{$w_i$==0}
\STATE Flag $\Leftarrow$ 1
\ENDIF
\STATE  $i \Leftarrow i-1$
\ENDWHILE

\STATE $//$ From LSB to $w_i$ do the M-CSD.
\WHILE{j \textless i}
\IF{$w_{j+4}...w_j == 11011$}
\STATE $w_{j+2}w_{j+1}w_j \Leftarrow 10\overline{1}$
\STATE $j \Leftarrow j+2$
\ELSIF{$w_{j+4}...w_j == \Bar{1}\Bar{1}0\Bar{1}\Bar{1}$}
\STATE $w_{j+2}w_{j+1}w_j \Leftarrow \overline{1}01$
\STATE $j \Leftarrow j+2$
\ELSIF{ $w_{j+2}w_{j+1}w_j==111$}
\STATE $k \Leftarrow j+2$
\WHILE{$w_k==1$}
\STATE $k \Leftarrow k+1$
\ENDWHILE
\STATE $w_{k}w_{k-1}...w_j \Leftarrow 10...\overline{1}$
\STATE $j \Leftarrow k$
\ELSIF{ $w_{j+2}w_{j+1}w_j==\Bar{1}\Bar{1}\Bar{1}$}
\STATE $k \Leftarrow j+2$
\WHILE{$w_k==\Bar{1}$}
\STATE $k \Leftarrow k+1$
\ENDWHILE
\STATE $w_{k}w_{k-1}...w_j \Leftarrow \overline{1}0...1$
\STATE $j \Leftarrow k$
\ELSE
\STATE $j \Leftarrow j+1$
\ENDIF
\ENDWHILE

\RETURN W

\end{algorithmic}
\end{algorithm}



\begin{figure*}
 \centering
 \includegraphics[width=1\textwidth]{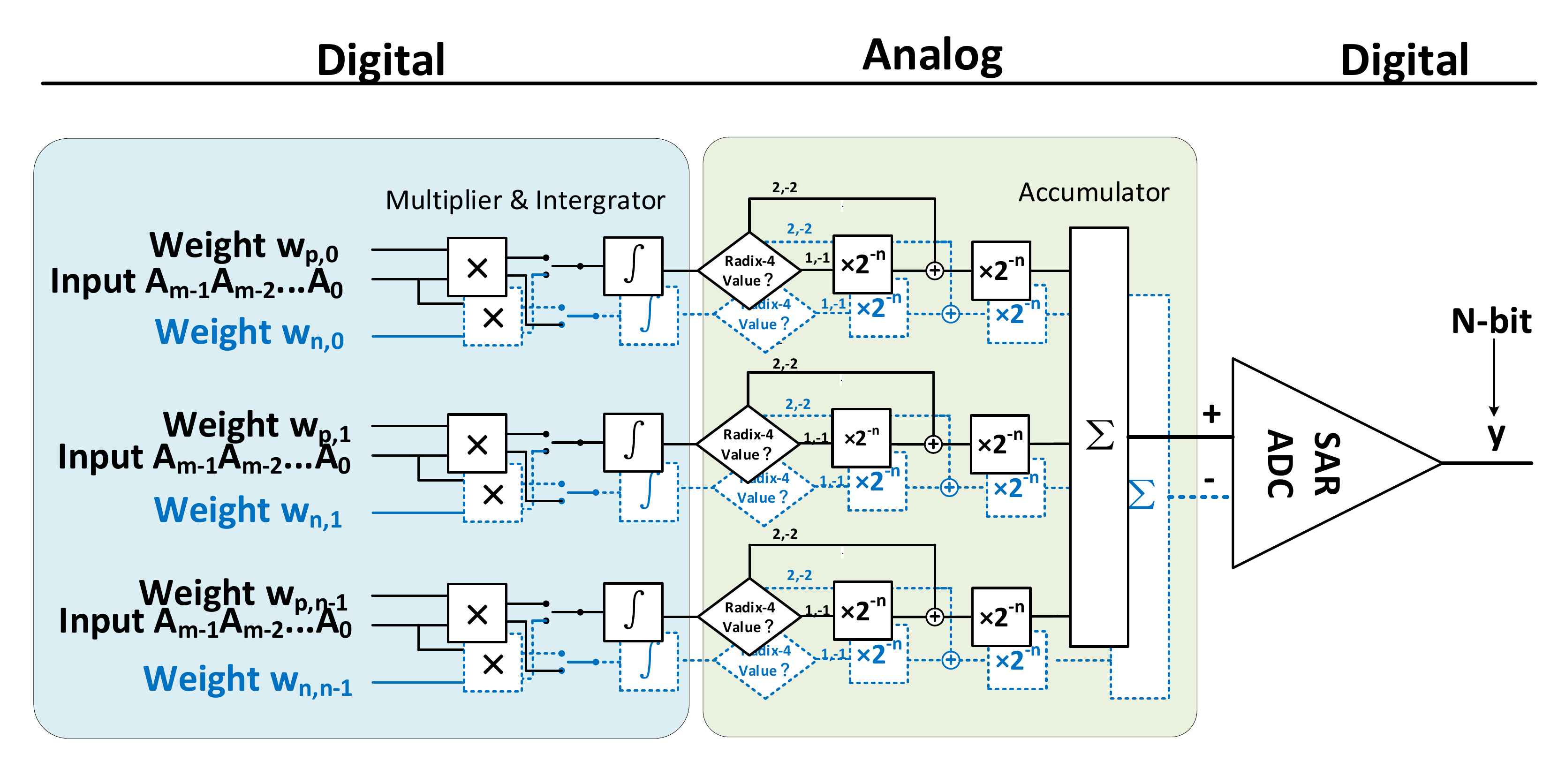}
 \caption{The block diagram of the integration scheme in the integral multiplier. It implements digital input/weight and analog MAC operations,
 and completes the analog-to-digital conversion output by a SAR ADC.}
 \label{Fig:inter_block}
\end{figure*}

\begin{figure*}
	\centering
	\subfigure[]{
	\centering
    \begin{minipage}[b]{\textwidth}
	\includegraphics[width=1\textwidth]{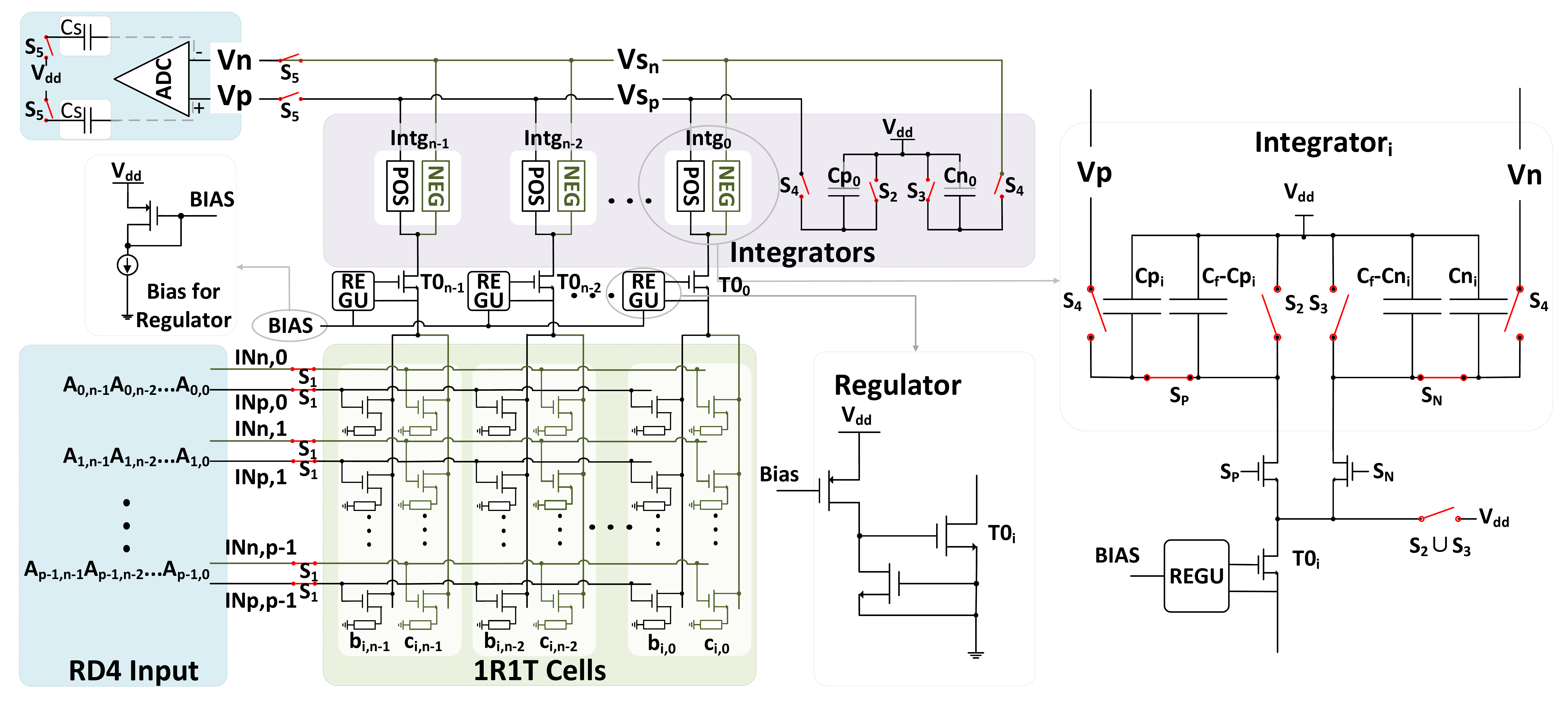}
	\label{Fig:integrator}
	\end{minipage}
	}
    \subfigure[]{
    \centering
    \begin{minipage}[b]{\textwidth}
    \includegraphics[width=1\textwidth]{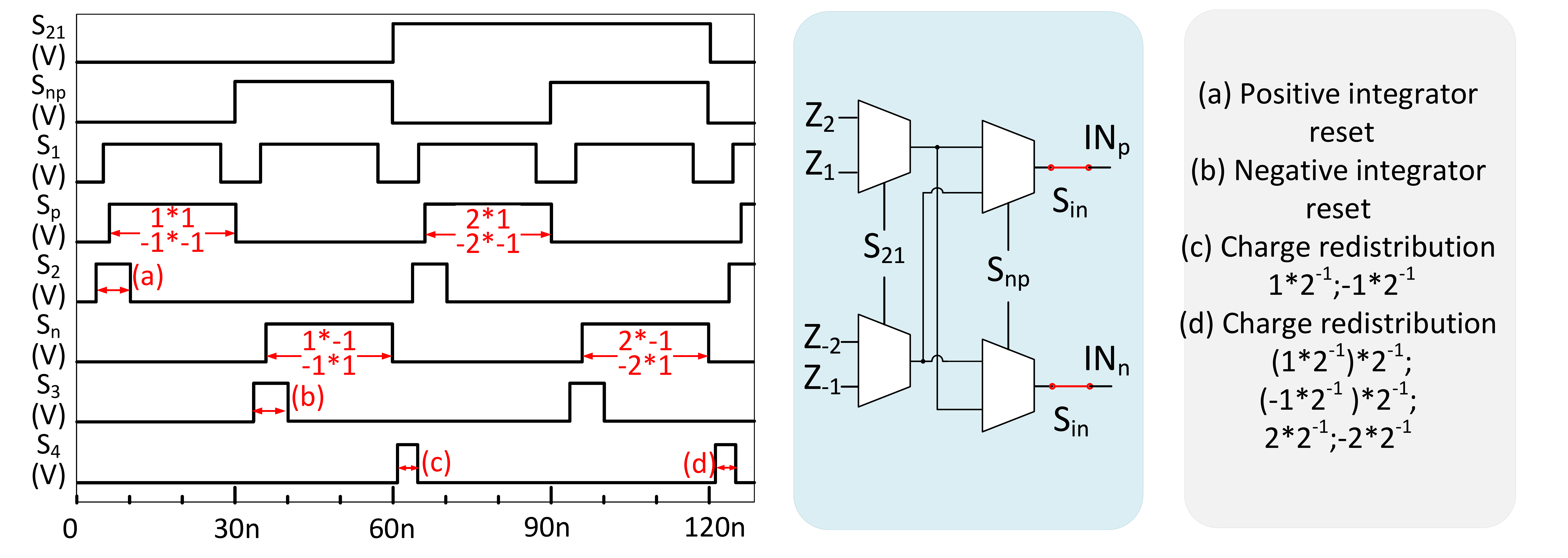}
    \label{Fig:timing}
    \end{minipage}
    }
    \subfigure[]{ 
	\centering
	\begin{minipage}[b]{\textwidth}
	\includegraphics[width=1\textwidth]{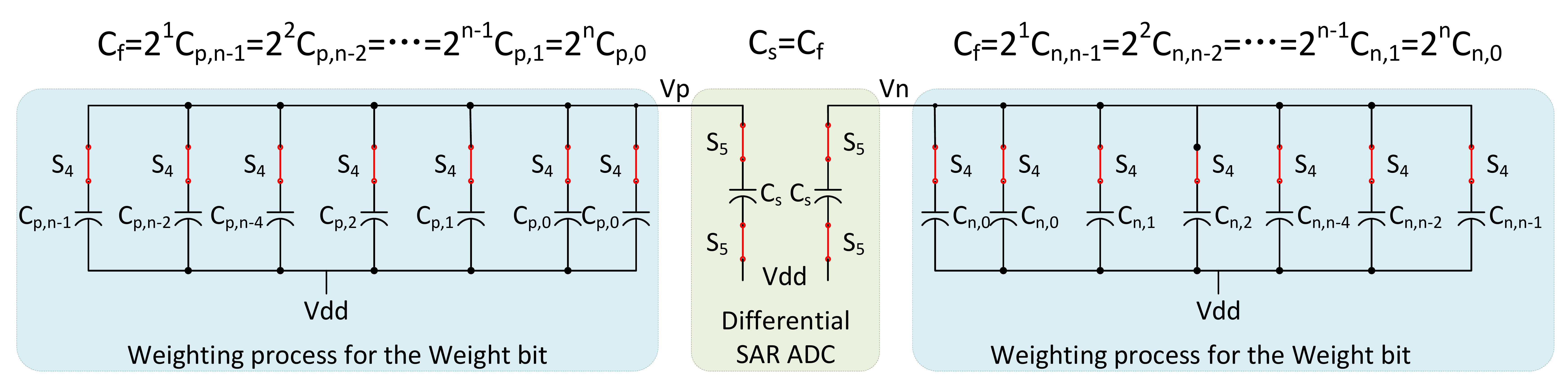}
    \label{Fig:weight-for-weight}
    \end{minipage}
    } 
    \caption{The proposed circuit schematics. (a) Integral multiplier with a symmetrical structure. The regulators are used to maintain the drain voltage of the 1R1T cells. All regulators
     have the same bias current. (b) The control logic of 1-bit input data. (c) The weighting process  for the weight bit and the input data are completed simultaneously in the charge redistribution phase. }
\end{figure*}

\subsection{Neuron Circuit}
The integral multiplier in the proposed CIM core is designed for massive parallel MAC operations and data transmission from digital to analog.
\cite{zhang2020robust} uses operational amplifiers to finish the integral operation. However, the static power consumption of the amplifier is not optimized for low power computing. 
Therefore, regulated passive neuron taken from \cite{zhang2020regulator} is adopted in our scheme to propose a differential passive integrator. As shown in Fig. \ref{Fig:inter_block}, 
digital inputs and digital weights are differentially multiplied and accumulated at the neurons. The proposed integration scheme contains three phases: positive integration, negative 
integration, and charge redistribution. The integration phases are used to  perform non-weighted MAC operations for inputs and weights. Therefore, each integrator has the same integral
voltage for different input bits and weight bits. The charge redistribution phase is used to perform the weighting process for M-RD4 digits ($4^{0} , 4^{1} , ... ,4^{m-1}$ from LSB to 
MSB, where m is the length of M-RD4 code, and $m=\left \lceil \frac{n}{2} \right\rceil$).


The integral neuron is designed as a symmetrical structure complete the positive and negative MAC operations separately. The differential integrator is illustrated in 
Fig. \ref{Fig:integrator}. The M-RD4 inputs are sequentially sent to the word lines from LSB to MSB. The RRAM model used in the 1R1T cells is around 10 G$\Omega$ in High Resistance State
 (HRS) and 10 M $\Omega$ in Low Resistance State (LRS) \cite{rram1,rram2}. The 1R1T cells are used in  pairs to store $b_i$ and $c_i$ mentioned in Eq  (\ref{EQ:diff}). The positive circuit
 is used for MAC operations whose results are positive ($I_p\times W_p + I_n \times W_n$), while the negative circuit is used for MAC operations with negative results 
 ($I_p \times W_n + I_n \times W_p$), where $I_p$, $I_n$, $W_p$, and $W_n$ are the positive input, negative input, positive weight, and  negative weight, respectively. In this way, the 
 number of the discharge path is reduced. What's more, the positive and negative circuits are compensated to each other, effectively reducing the influence of parasitic parameters. 
 Therefore, the proposed integrator can achieve higher accuracy with lower power.  $S_{1}$ controls the data input, $S_P$ controls the positive integral operation, and $S_N$ controls 
 the negative operation. $S_2$, $S_3$, and $S_4$ control the integration phase and the charge redistribution phase. $S_5$ controls the sample phase and the conversion phase of the ADC.

During the positive integration phase, $S_4$ is open to separate each integrator.  After that, $S_2$ and $S_P$ are closed to clear the charge in positive integral capacitors. Then $S_1$
is closed to input the M-RD4 data ($IN_p=1$, and$IN_n=-1$), and $S_2$ is open to complete the 1-bit MAC of $1\times1+ -1\times1$. After the positive integration phase,  $S_P$ is 
open to keep the charge in $Cp_i$, and $S_1$ is open to ensure no power is consumed by the 1R1T cells. During the negative integration phase, $S_4$ is still open to make sure the 
integrator are separated. $S_3$ and $S_N$ are closed to clear the charge in negative capacitors. After that, $S_1$ is closed with the input $IN_p=-1$, and  $IN_n=1$. The phase complete 
the 1-bit MAC of $-1\times1+ 1\times-1$. After two integration phases, $S_4$ and $S_5$ are closed  to complete the charge redistribution phase, where the equivalent analog voltage 
($V_p$ for positive and $V_n$ for negative) is generated. According to the derivation process of \cite{zhang2020regulator}, the positive or negative integration voltage after one step 
of the charge redistribution phase is

\begin{equation}
    \begin{split}
      V_S &= V_S^--k(2^{-1}\sum_{i=0}^{p-1}A_iG_{i,n-1} +2^{-2}\sum_{i=0}^{p-1}A_iG_{i,n-2}+...\\
      &+2^{-n+1}\sum_{i=0}^{p-1}A_iG_{i,0})
    \end{split}
  \end{equation} 
where $V_S$ represents $V_{Sp}$ or $V_{Sn}$, and $V_S^-$ represents the initial integral voltage. $k = \frac{V_BT}{C_f} $, $p$ is the number of the input layers, $A_i$ is 1-bit M-RD4 input of the $i_{th}$ input line, and $T$ is the fixed time period for each integration. $G_i$ is the conductance of each binary-RRAM cell, which is $1/R_H$ and $1/R_L$ when it is in the HRS and LRS, respectively. 

In the proposed scheme, the input pulse has only two possible values, which can effectively reduce the 1R1T cells' reading variation. Therefore, 1-bit M-RD4 data with different values are computed sequentially. The bits in M-RD4 have the relationship $ A_{i,m-1}=4^1A_{i,m-2}=...4^{m-1}A_{i,0}$, which means each bit needs two steps of charge redistribution operation to achieve the weighting process for input data.   As shown in Fig. \ref{Fig:timing}, four integration phases (two positive and two negative) and two charge redistribution phases are needed to complete the computing and weighting process for 1-bit M-RD4 data.  The first two integration phases mentioned above compute the layers whose input is `1' or `-1'. As shown in Fig. \ref{Fig:weight-for-weight}, the first charge redistribution phase uses the sampling capacitor $C_S$ to complete the weighting process for input data. Let $C_S=C_f$, the charge on the capacitors $C_{n-1}C_{n-2}...C_0$ and $C_S$ is equally divided after the charge redistribution operation. Taking the positive integrator as an example, the voltage $V_{p,a}$ of $C_S$ can be expressed as
\begin{equation}
   V_{p,a}=\frac{1}{2}(V_{Sp,a}+V_{p}^-)
\end{equation}
where $V_{p}^-$ represents the previous positive voltage in $C_S$, $V_{Sp,a}$ represents  the positive integration voltage for layers with input `1' and `-1'. In the second two integration phases, the layers with input `2' or `-2' are input and computed. The positive voltage of $C_S$ after the second charge redistribution phase is 
\begin{equation}
    V_{p}=\frac{1}{2}(V_{Sp,b}+V_{p,a})=\frac{1}{2}V_{Sp,b}+\frac{1}{4}V_{Sp,a}+\frac{1}{4}V_{p}^-
\label{Eq:weight for input}
\end{equation}
where $V_{Sp,b}$ is the positive integration voltage for layers with input `2' and `-2', $V_{p}^-$ is the positive output voltage after the last input bit is computed. Eq (\ref{Eq:weight for input}) described the for loop process for each bit of the input data. Therefore the input data is weighted by $4^m-1,4^m-2,...,4^0$ from LSB to MSB. Initially $V_{p}$ is reset to Vdd. After m-bit input data are computed, it can be expressed as 
\begin{equation}
 V_{p}=4^{-m}V_{dd}+4^{-m}V_{Sp,0}+...+4^{-1}V_{Sp,m-1},\\
\end{equation}
where $V_{Sp,i} = V_{Sp,a,i}+2V_{Sp,b,i}$, the change of the $V_p$ is 
\begin{equation}
        \Delta V_{p} = V_{dd} - V_{p}
        =4^{-m}\sum_{i=0}^{m-1}4^i\Delta V_{Sp,i}
\end{equation}
where $\Delta V_{Sp,i} = \Delta V_{Sp,a,i} +2\Delta V_{Sp,b,i}$, $V_{Sp,i}$ is the $i_{th}$ positive integration voltage, and $\Delta V_{Sp,i}$ is the change of $V_{Sp}$ in the $i_{th}$ integration.
Therefore, the output voltage is
\begin{equation}
        V_{out}=\Delta V_p - \Delta V_n \\
      =4^{-m}\sum_{i=0}^{m-1}4^i(\Delta V_{Sp,i}-\Delta V_{Sn,i})
\end{equation}

\subsection{Mapping}

There are several methods to implement the convolution layers and fully connected layers on cross-point arrays \cite{gao2016demonstration, wang2020deep, cai2019fully}. To estimate the network level energy efficiency of the proposed scheme, the mapping method in \cite{zhang2020robust} is adopted. Both convolution kernel in convolution layers and weight matrix in fully connected layers are mapped into the cores.  A convolution kernel whose size is $C_{in}*k*k*C_{out}$ is firstly transform it to a 2D matrix with size $(C_{in}*k*k)\times C_{out}$. The proposed scheme has a cross-point array size of $256\times512$, and can implement a $256\times256$ matrix. Therefore, the number cores to implement the kernel is $\left \lceil \frac{C_{in}*k*k}{256} \right \rceil \times \left \lceil \frac{C_{out}}{256} \right \rceil$. The adders are integrated in the router unit to sum the results of different cores if the kernel size is larger than 256. For an $M*N$ fully connected layer, the weight matrix can be mapped into $\left \lceil \frac{M}{256} \right \rceil \times \left \lceil \frac{N}{256} \right \rceil$ cores, respectly.

\section{Simulation Results}
In this section, both circuit-level and network-level evaluation results are provided.  The circuit-level simulation verifies the circuit's functionalities and shows the  energy and accuracy benefits of the proposed core. The network-level evaluation presents the performance comparison with other related works. The circuit-level simulations are done in Cadence Analog Mixed Signal (AMS) with a 45nm generic Process Design Kit (PDK). The RRAM model proposed by \cite{jiang2016compact} is adopted in the circuit simulations. The network-level simulations are done on the PyTorch platform.

\subsection{Functionality}

\begin{figure}[h]
  \centering
  \includegraphics[width=0.45\textwidth]{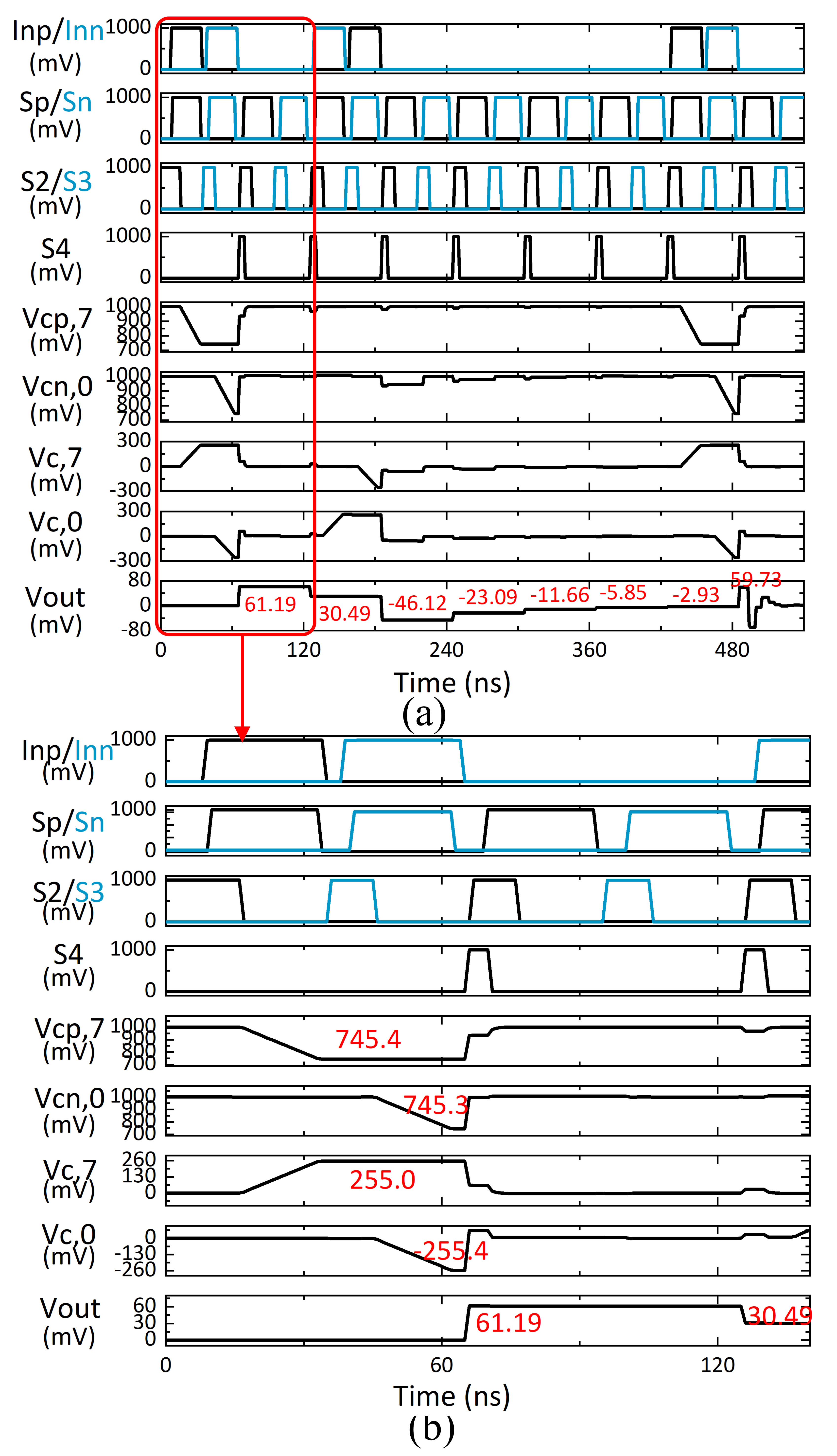}
  \caption{The transient simulation results of (a) the computing progress for 1-bit M-RD4 input, and (b) the whole MAC operation for M-RD4 input `2,0,-1,1' and differential  weight 8'b10000000-8'b00000101.}
  \label{Fig:8b_tran}
\end{figure}

The transient simulation is performed to verify the correct function of the circuit. A random input 125 (binary representation: 8'b01111101, M-RD4 representation: 2, 0, -1, 1) is sent to the CIM core to complete the MAC operation with a random weight 123 (binary representation: 8'b01111011, differential representation:  8'b10000000-8'b00000101). Fig.\ref{Fig:8b_tran} (a) shows the whole MAC operations. The input bits is computed from LSB to MSB. From 0 ns to 130 ns, the circuit completes the MAC operation for the M-RD4 bit `1'. As shown in Fig. \ref{Fig:8b_tran} (b), $Vc_{p,7}$ is the integration voltage of the positive capacitor $C_{p,7}$, which is reset to 1 V when $S_2$ is closed. From 16 ns to 31 ns, $S_P$ is closed and $Vc_{p,7}$ is decreased to 745.4 mV linearly to complete the multiplication of $1\times1$. $Vc_{n,0}$ is the  integration voltage of the negative capacitor $C_{n,0}$, which is reset when $S_3$ is closed. The multiplication of $1\times-1$ is completed from 47 ns to 62 ns where $Vc_{n,0}$ is decreased to  745.3 mV linearly. From 64 ns to 70 ns, $S_4$ is closed to complete the charge redistribution phase, and the output voltage $V_{out}$ is 61.19 mV. From 66 ns to 124 ns, $Vc_{p,7}$ and $Vc_{n,0}$ are kept at 1 V since no data is input. After the  second charge redistribution phase, the output voltage $V_{out}$ is halved to 30.49 mV. The computing of the M-RD4 input `1' is completed. Using the difference as output can effectively reduce the impact of parasitic parameters on accuracy. After 8 cycles of integration and charge redistribution phase, the output voltage $V_{out}$ is 59.73 mV. The digital result is 8'b00111011. The theoretical results are 59.89 mV and 8'b00111011, respectively. Therefore, the proposed scheme achieves its design requirement.

\subsection{Robustness Analysis}

\begin{figure*}[htpb]
  \centering
  \subfigure[Input Value]{
    \centering
    \begin{minipage}[b]{0.28\textwidth}
      \includegraphics[width=1\textwidth]{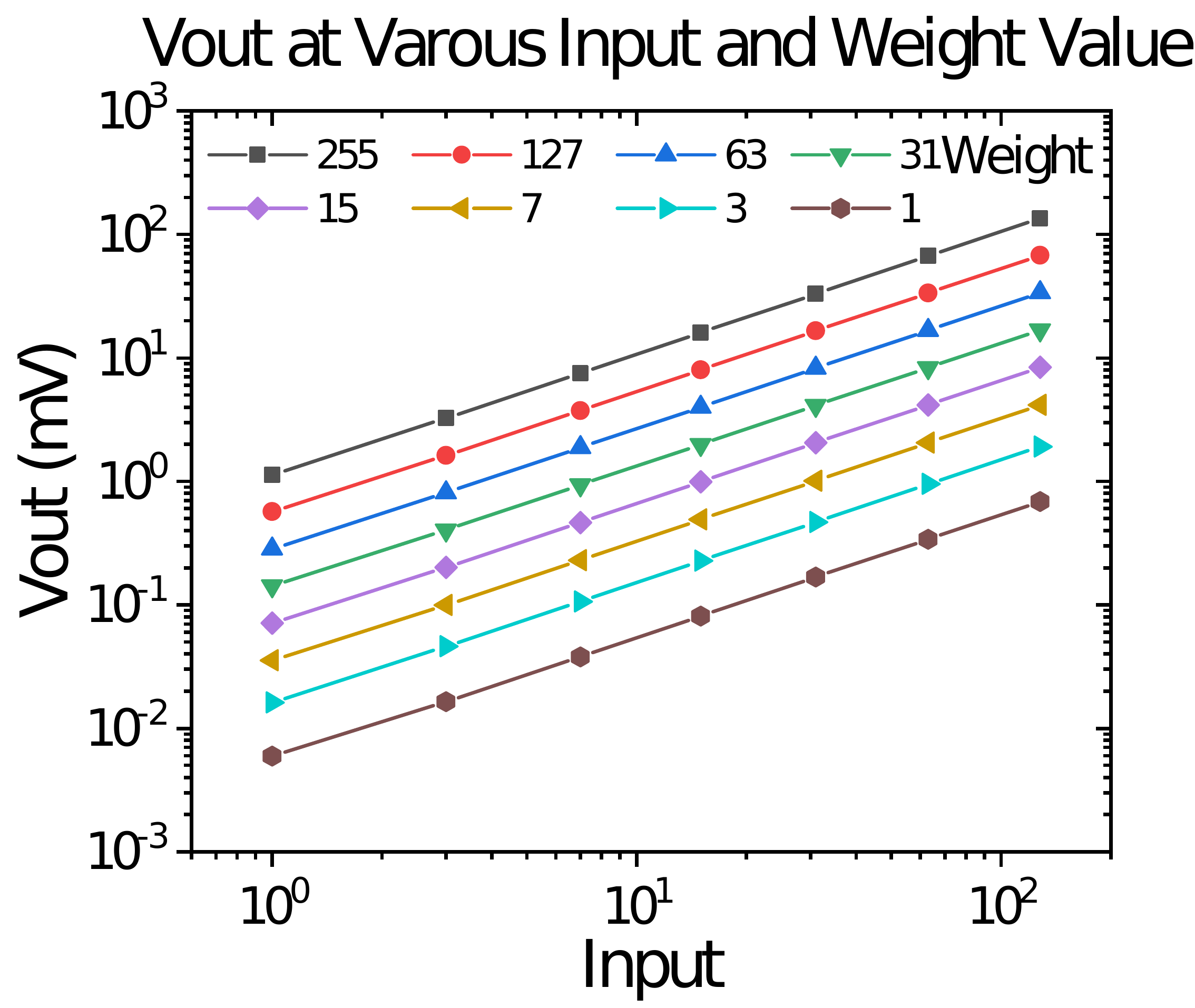}
    \end{minipage}
    \label{lin_value_input}
  }\hspace{0.05\textwidth}
  \subfigure[Input Lines]{
    \centering
    \begin{minipage}[b]{0.28\textwidth}
      \includegraphics[width=1\textwidth]{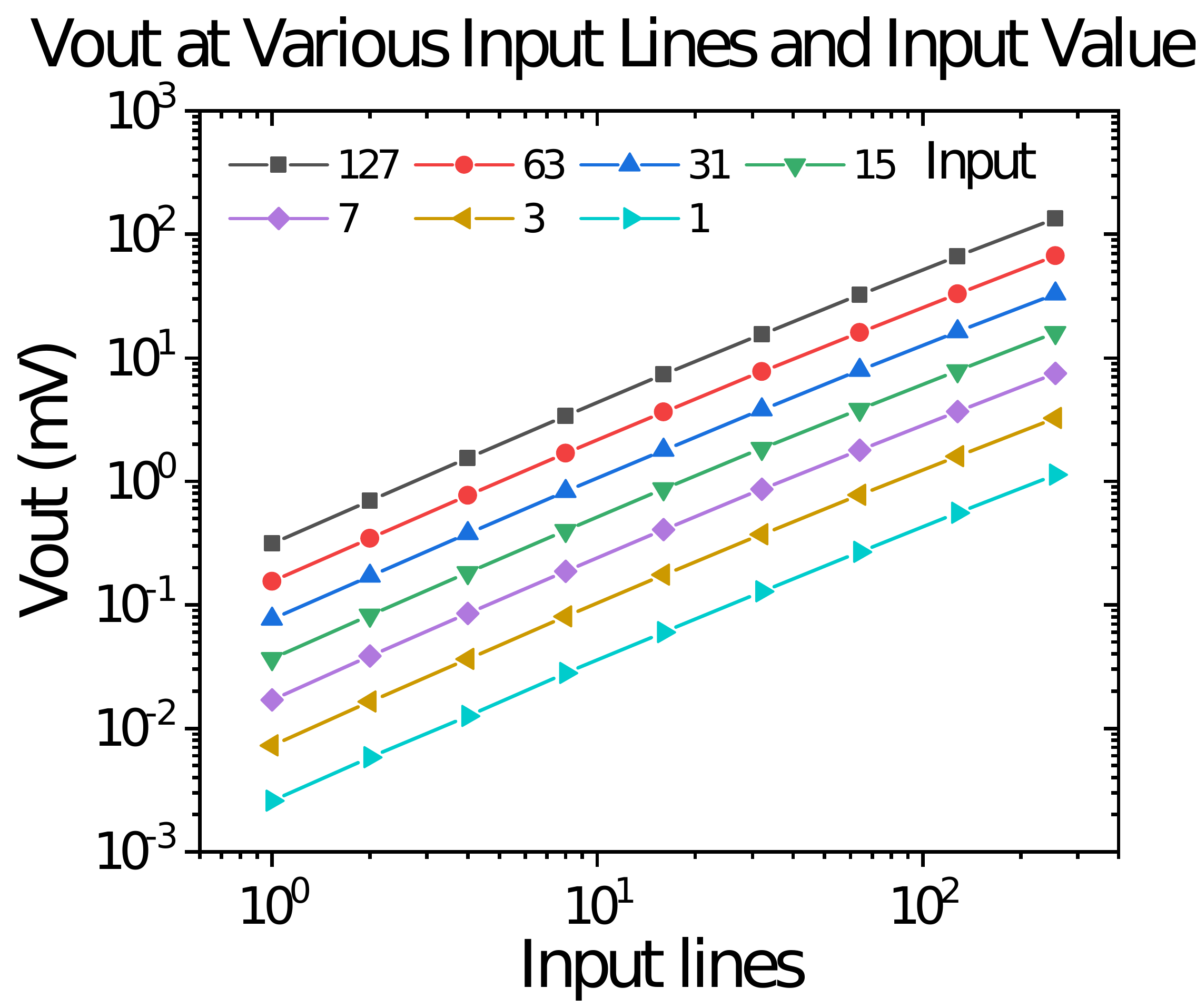}
    \end{minipage}
    \label{lin_lines_input}
  }\hspace{0.05\textwidth}
  \subfigure[Weight]{
    \centering
    \begin{minipage}[b]{0.28\textwidth}
      \includegraphics[width=1\textwidth]{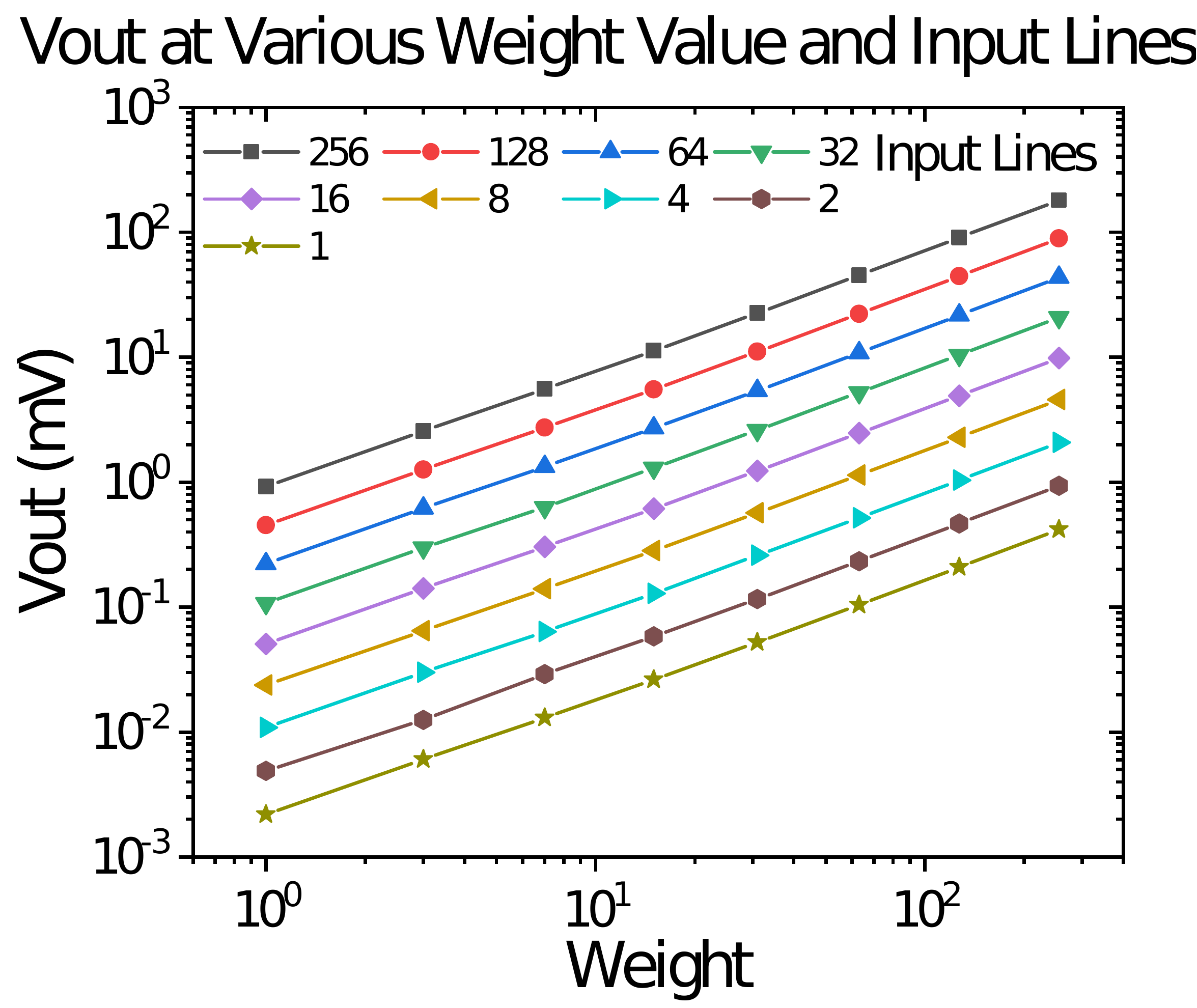}
    \end{minipage}
    \label{lin_value_weight}
  }
  \caption{The output voltage $V_{out}$ at various (a)input data and RRAM weights, (b) input lines and input data, (c) RRAM weights and input lines in the proposed scheme} 
  \label{Fig:linearity}
\end{figure*}

\begin{figure*}[htpb]
  \centering
  \subfigure[]{
    \centering
    \begin{minipage}[b]{0.28\textwidth}
      \includegraphics[width=1\textwidth]{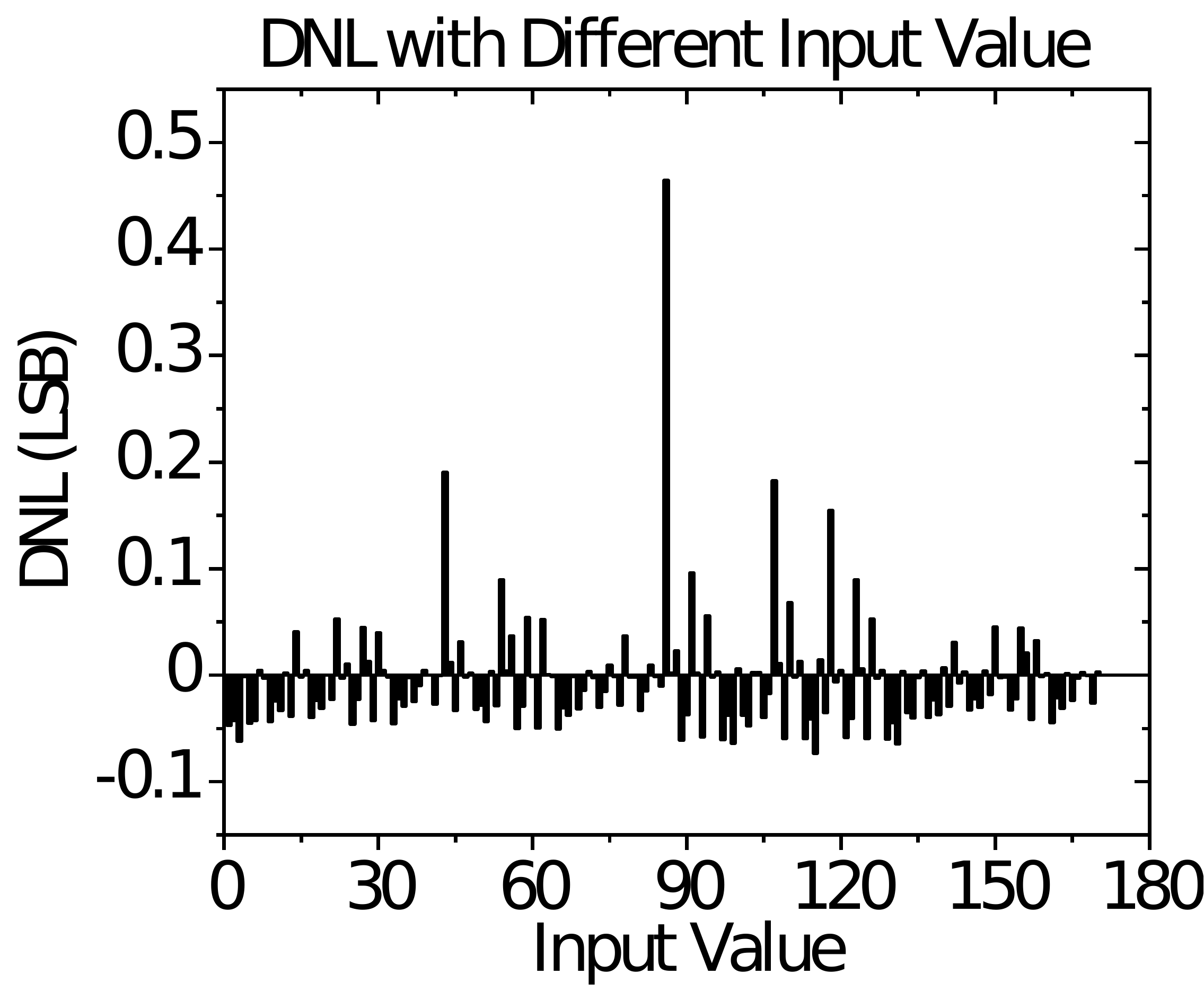}
    \end{minipage}
    \label{DNL_INL_1}
  }\hspace{0.05\textwidth}
  \subfigure[]{
    \centering
    \begin{minipage}[b]{0.28\textwidth}
      \includegraphics[width=1\textwidth]{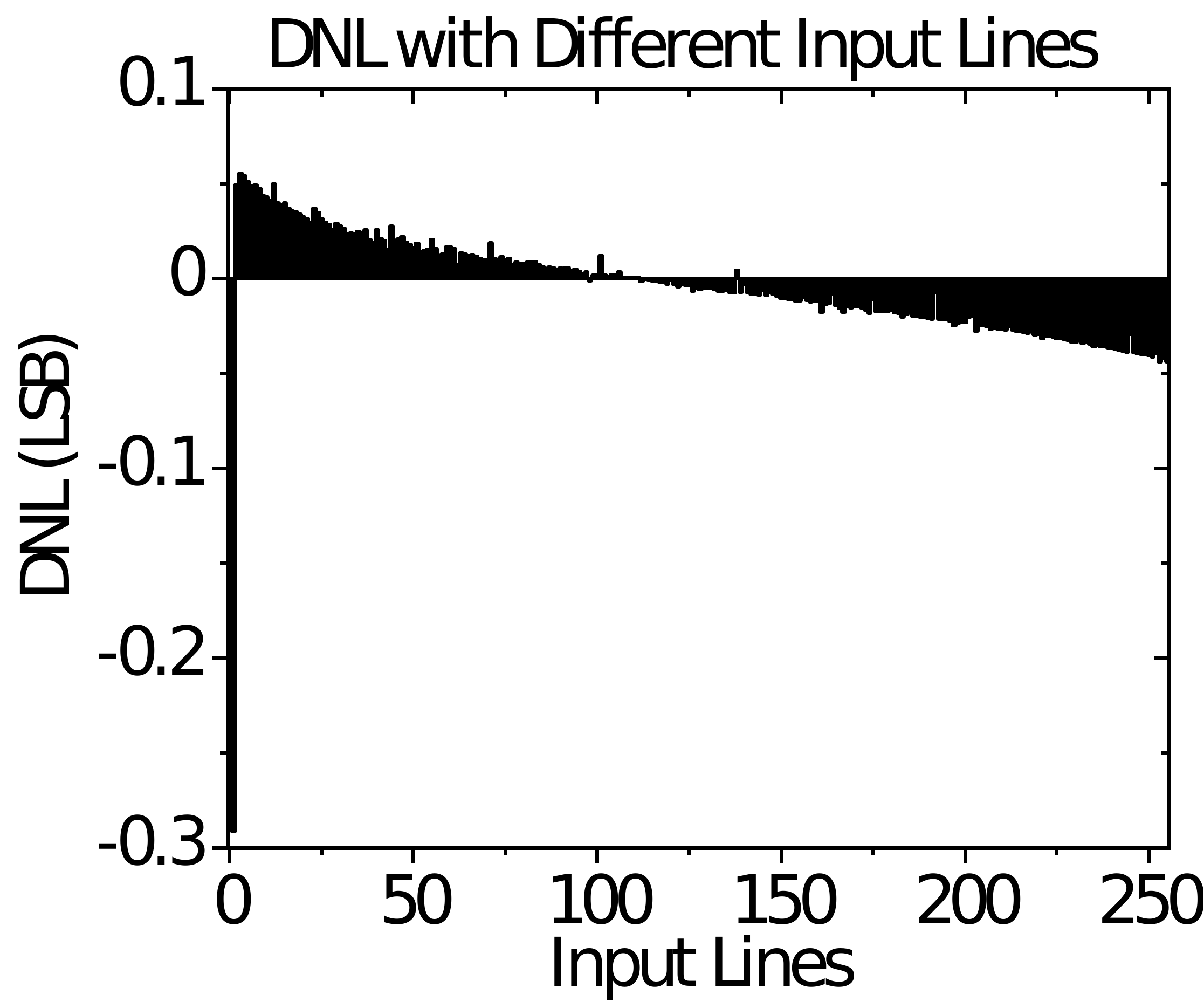}
    \end{minipage}
    \label{DNL_INL_2}
  }\hspace{0.05\textwidth}
  \subfigure[]{
    \centering
    \begin{minipage}[b]{0.28\textwidth}
      \includegraphics[width=1\textwidth]{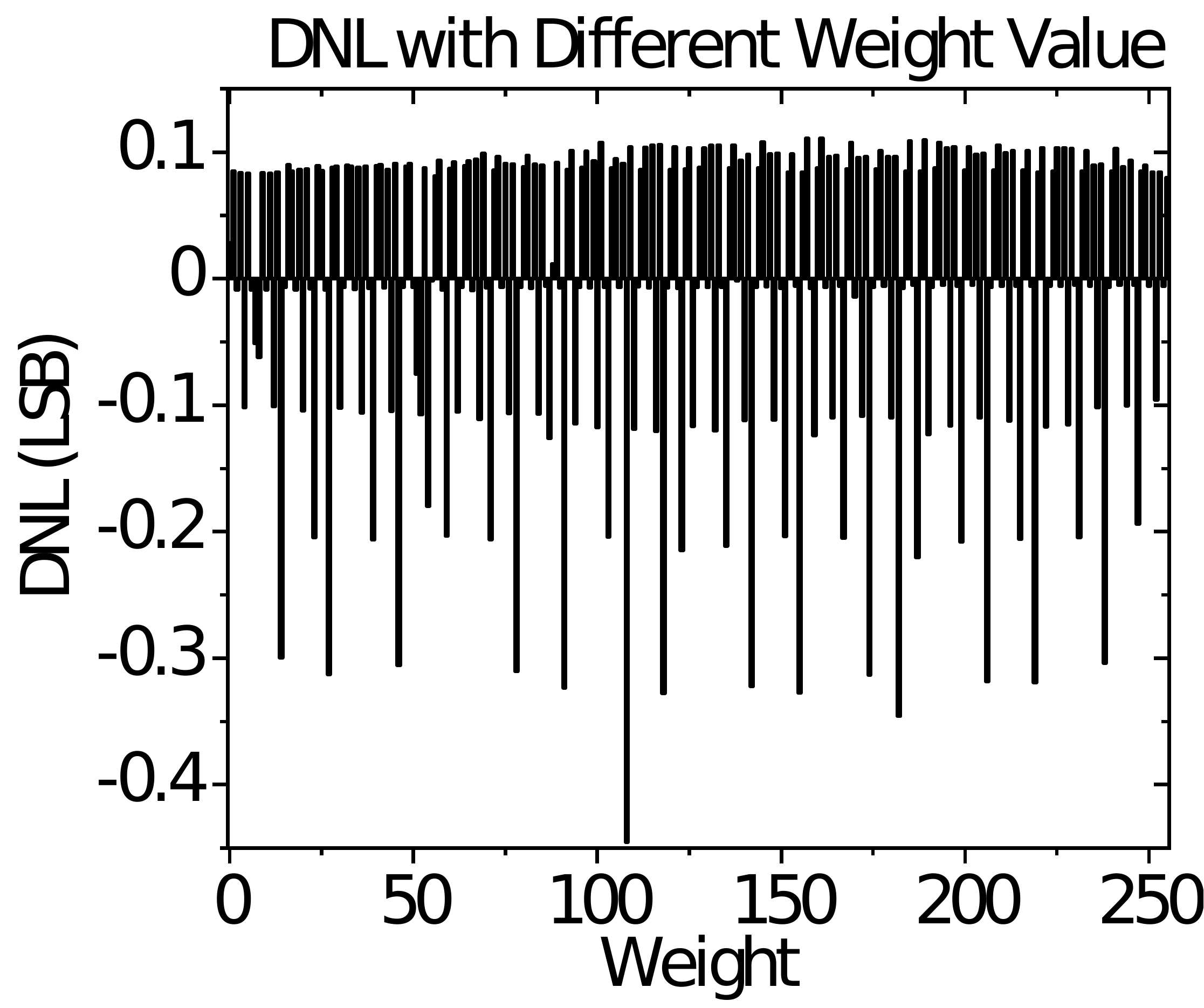}
    \end{minipage}
    \label{DNL_INL_3}
  }
  \subfigure[]{
    \centering
    \begin{minipage}[b]{0.28\textwidth}
      \includegraphics[width=1\textwidth]{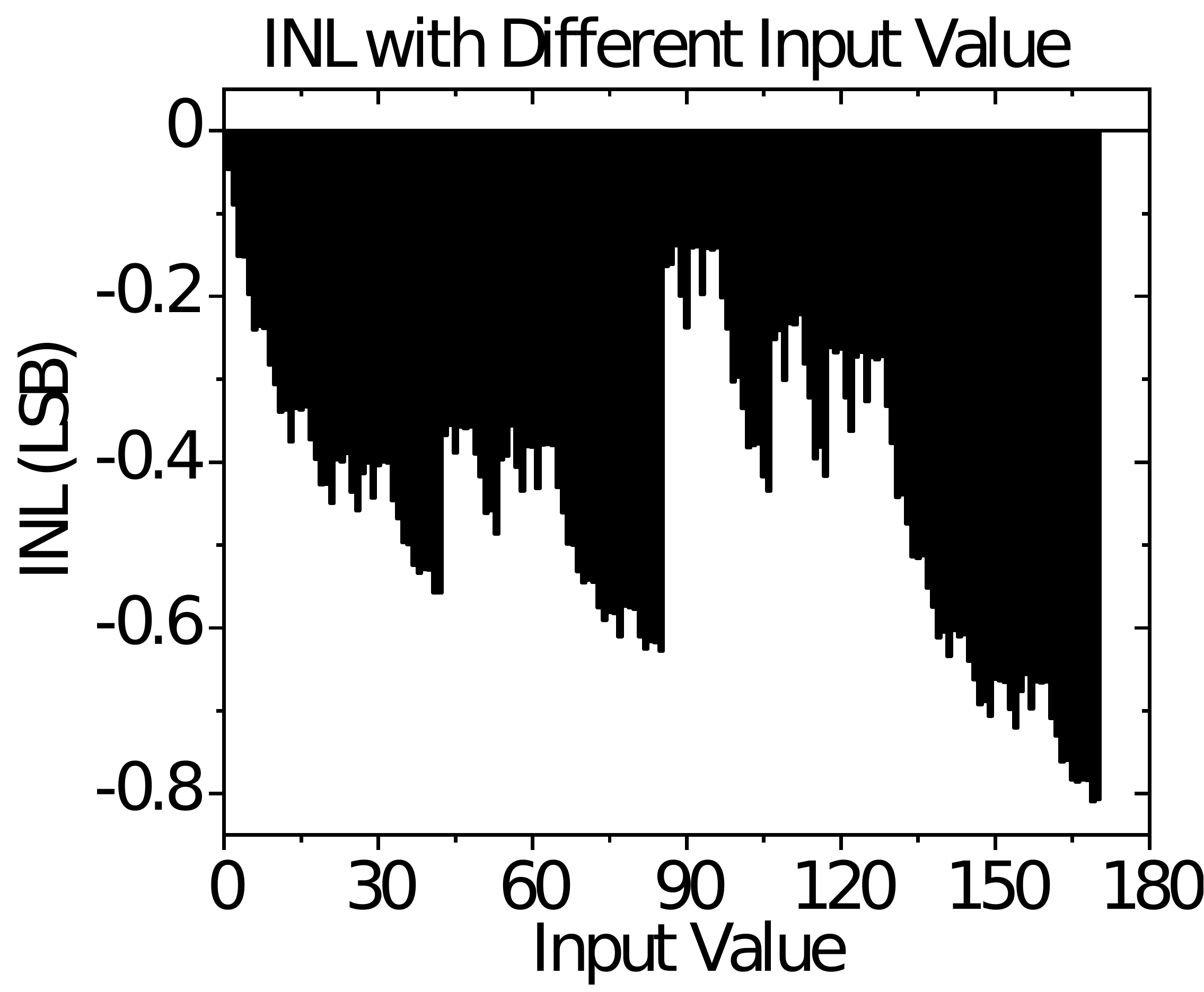}
    \end{minipage}
    \label{DNL_INL_4}
  }\hspace{0.05\textwidth}
  \subfigure[]{
    \centering
    \begin{minipage}[b]{0.28\textwidth}
      \includegraphics[width=1\textwidth]{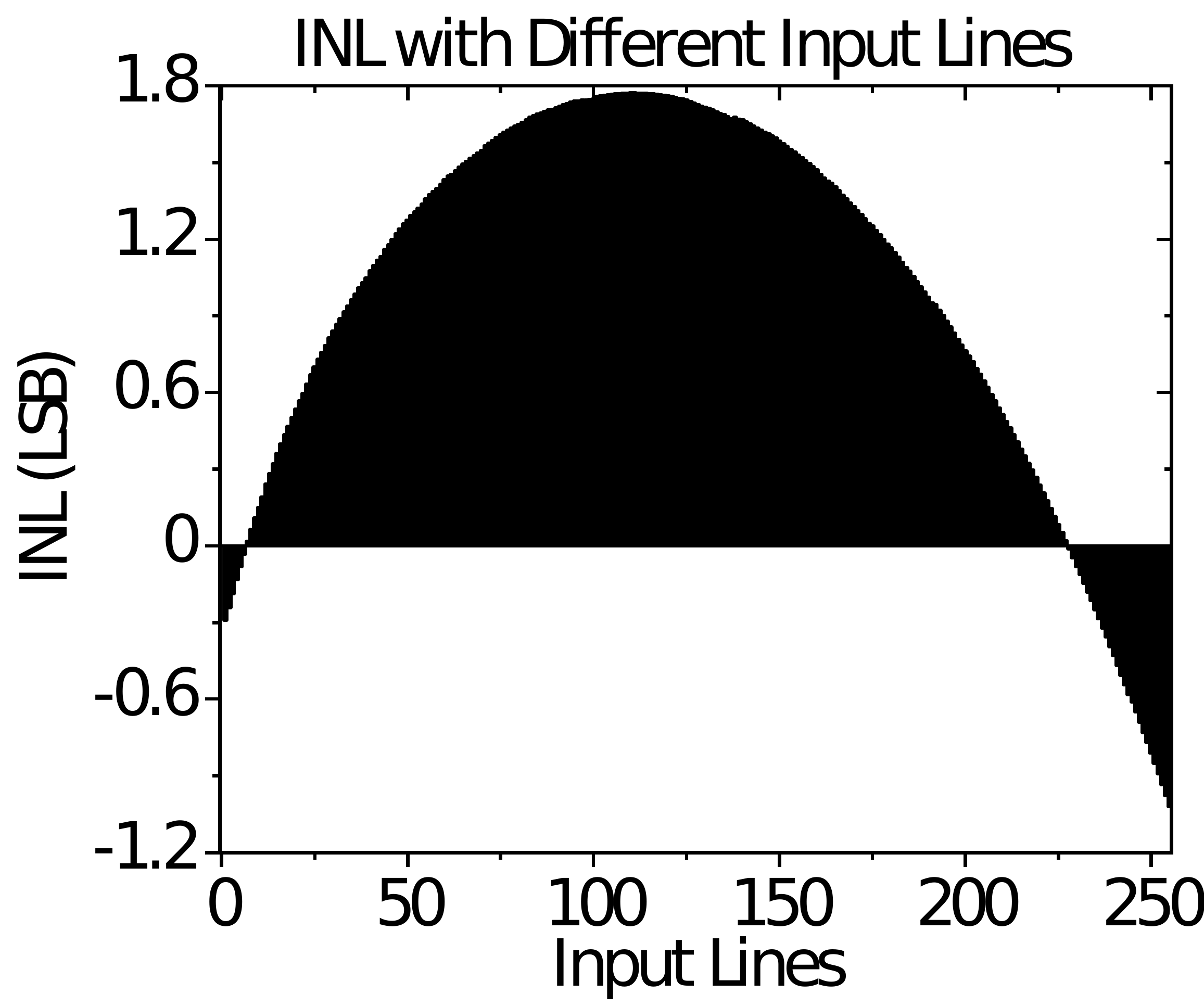}
    \end{minipage}
    \label{DNL_INL_5}
  }\hspace{0.05\textwidth}
  \subfigure[]{
    \centering
    \begin{minipage}[b]{0.28\textwidth}
      \includegraphics[width=1\textwidth]{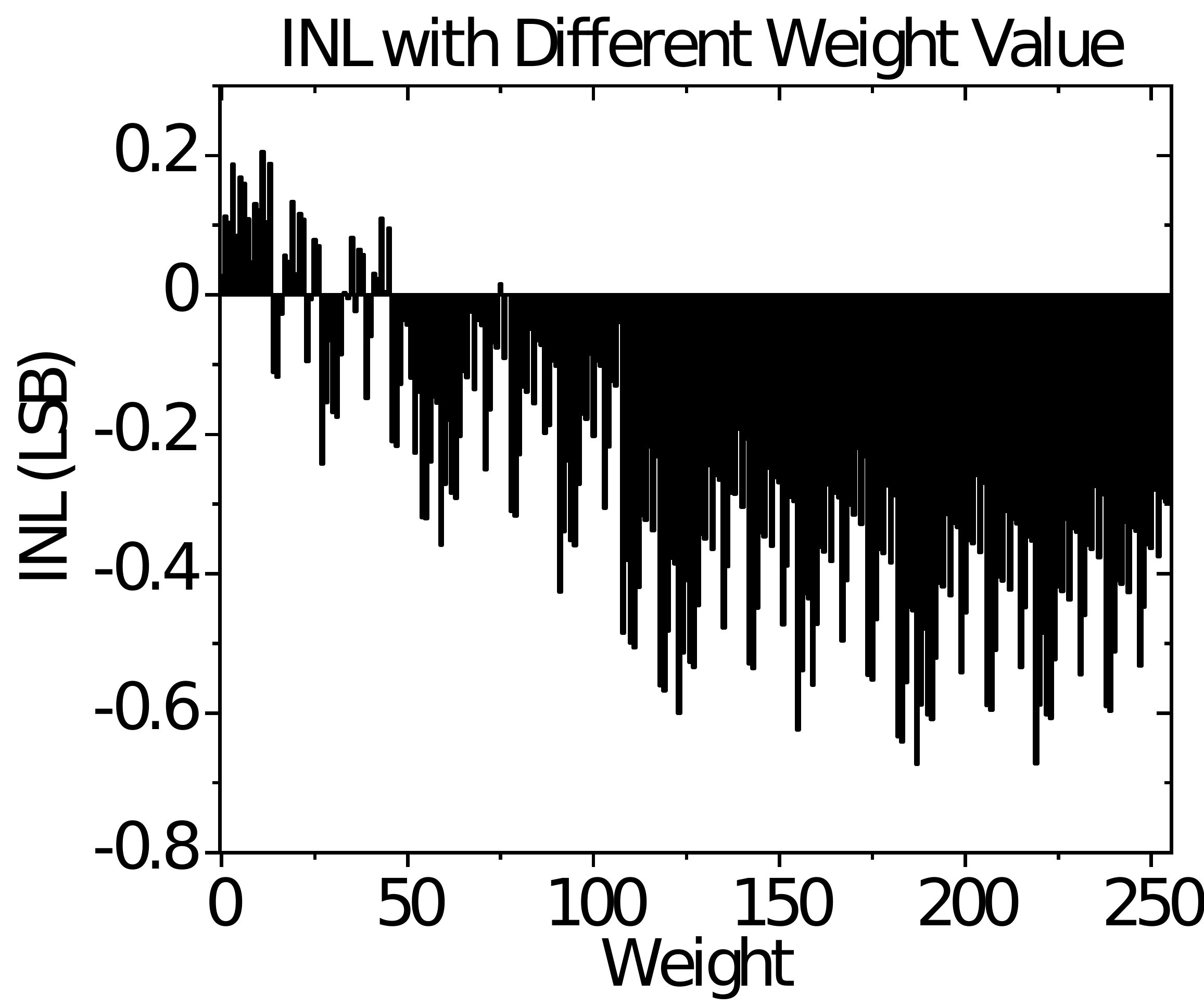}
    \end{minipage}
    \label{DNL_INL_6}
    }\hspace{0.05\textwidth}
  \caption{The simulated DNL in terms of (a)input value, (b)input lines, (c) weight value and INL in terms of (d)input value, (e)input lines, (f)weight value in the proposed scheme} 
  \label{Fig:DNL_INL}
\end{figure*}
Fig. \ref{Fig:linearity} shows the relationship between analog output  $V_{out}$ ($\Delta V_p - \Delta V_n$) and (a) the digital input, (b) the number of input lines, and (c) the digital weight. The results show that the proposed scheme achieves high linearity and accuracy. Fig. \ref{Fig:DNL_INL} (a), (b), and (c) show the Differential Non-linearity (DNL) of the proposed scheme with different (a) input value, (b) input lines, and (c) weight value, respectively. Fig. \ref{Fig:DNL_INL} (d),(e), and (f) show the Integration Non-Linearity (INL), respectively. The simulated DNLs (INLs) in terms of the digital input, the digital weight, and the number of input lines are +0.464/-0.073 LSB (-0.047/-0.809 LSB), +0.055/-0.291 LSB (+1.772/-1.061 LSB), and +0.111/-0.445 LSB (0.205/-0.673 LSB).

\begin{table}[htpb]
    \centering
    \caption{PVT simulation on ENOB}
    \begin{tabular}{|c|c|c|c|c|c|c|}
    \hline
    \multicolumn{2}{|c|}{Process}&\multicolumn{2}{c|}{ff}&\multicolumn{2}{c|}{ss}&tt  \\
    \hline
    \multicolumn{2}{|c|}{Temprature ($^{\circ}$C)}&-40&80&-40&80&27\\
    \hline
    \multirow{3}{*}{Voltage (V)}&1& & & & & 7.42\\
    \cline{2-7}
    &0.9&7.41&7.28&7.29&7.21&\\
    \cline{2-7}
    &1.1&7.11&7.16&7.25&7.11&\\
    \hline
    \end{tabular}
    \label{Table:PVT}
\end{table}

 Different process, voltage and temperature are chosen to do the PVT simulation to verify the robustness of the circuit. ENOBs, as shown in Table \ref{Table:PVT}, are all greater than 7.1 bits in different PVT combinations. Therefore, the proposed scheme is reliable with different variations of the process, voltage, and temperature.

\begin{table}[htpb]
    \centering
    \caption{CIM core performance comparison between MBRAI, RPN\&BLM and the proposed}
    \resizebox{0.45\textwidth}{14mm}{
    \begin{tabular}{|c|c|c|c|c|}
    \hline
               & MBRAI \cite{zhang2020robust}&RPN\&BLM \cite{zhang2020regulator}&\multicolumn{2}{c|}{Proposed} \\ \hline
       \multirow{2}{*}{Supply}  & \multirow{2}{*}{1.1 V} & \multirow{2}{*}{1 V} & M-RD4 Recoder & 0.6 V      \\ 
       \cline{4-5}
       &&&Neuron Circuit& 1 V \\
       \hline
       Computing speed & 1.85 M/s&1.85 M/s & \multicolumn{2}{c|}{1.85 M/s}\\ \hline
       SFDR & 67.42 dB &59.13 dB& \multicolumn{2}{c|}{63.41 dB} \\ \hline
       SNDR & 45.48 dB &46.13 dB& \multicolumn{2}{c|}{46.48 dB} \\ \hline
       ENOB & 7.26 bit &7.37 bit& \multicolumn{2}{c|}{7.42 bit} \\ \hline
    \end{tabular}
    }
    \label{Table:core_performance}
\end{table}
\begin{table}[htpb]
    \centering
    \caption{Energy cost comparison between the proposed core and others}
    \resizebox{0.45\textwidth}{16mm}{
    \begin{tabular}{|c|c|c|c|c|c|}
    \hline
    \multicolumn{2}{|c|}{}&MBRAI\cite{zhang2020robust}&MBHS-mCNN\cite{yao2020fully}&RPN\&BLM\cite{zhang2020regulator}&Proposed\\
    \hline
    \multicolumn{2}{|c|}{Technology}&45 nm & 65 nm & 45 nm & 45 nm\\
    \hline
    \multicolumn{2}{|c|}{Supply}&1.1 V&-&1 V&0.6/1 V  \\
    \hline
    \multicolumn{2}{|c|}{System Frequency}&16.7 MHz&20 MHz & 16.7 MHz & 16.7 MHz\\
    \hline
    \multicolumn{2}{|c|}{Core Size}&256*256&128*256&256*256&256*512\\ 
    \hline
    \multirow{4}{*}{Power}&Amplifier&0.22 mW &-&-&-\\ 
    \cline{2-6}
    &ADC&4.04 uW & 25.47 uW & 4.04 uW & 3.99 uW\\
    \cline{2-6}
    &Regulator&-&-&1.11 uW & 0.55 uW\\
    \cline{2-6}
    &Core &199.68 mW & 7.44 mW & 3.61 mW & 2.00 mW \\
    \hline
    \end{tabular}
    }
    \label{Table:energy_cost}
\end{table}
\begin{table*}[htpb]
    \centering
     \caption{Core-level comparison between the proposed core and others}
    \resizebox{\textwidth}{28mm}{
    \begin{tabular}{|c|c|c|c|c|c|c|}
    \hline
       \multirow{2}{*}{Structure} & \multirow{2}{*}{Technology}&\multirow{2}{*}{Crossbar-size}&\multirow{2}{*}{Weight/data bit}&Throughput  & Power & Efficiency   \\
       &&&&(GOPS)&(mW)&(TOPS/s/W)\\
       \hline
       \multirow{2}{*}{SINWP\cite{SINWP}} & \multirow{2}{*}{55 nm} & \multirow{2}{*}{256*512} & fixed-3/fixed-1 &-&-&53.17\\
       \cline{4-7}
       &&&fixed-3/fixed-2&-&-&21.9\\ 
       \hline
       \multirow{3}{*}{MBRAI\cite{zhang2020robust}}&\multirow{3}{*}{45 nm}&\multirow{3}{*}{256*256}&fixed-3/fixed-1&1524&19.6&77.76\\
       \cline{4-7}
       &&&fixed-3/fixed-2&1040&26.8&38.8\\
       \cline{4-7}
       &&&fixed-8/fixed-8&121.4&199.68&0.61\\
       \hline
       MBHS-mCNN\cite{yao2020fully}&65 nm&128*256&fixed-8/fixed-8&81.82&7.348&11.15\\
       \hline
       7nm SRAM Macro\cite{dong2020sram} & 7 nm & 4 K & fixed-4/fixed-4 & 186.2 & 1.06 & 175.5 \\
       \hline
       \multirow{3}{*}{RPN \& BLM\cite{zhang2020regulator}} & \multirow{3}{*}{45 nm} & \multirow{3}{*}{256*256} & fixed-2/fixed-2&1092.2&1.975&553.01\\
       \cline{4-7}
      &&&fixed-4/fixed-4&546.1&2.66&205.30\\
      \cline{4-7}
      &&&fixed-8/fixed-8&121.4&3.61&33.63\\
      \hline
      \multirow{2}{*}{Synapses Integrated Analog Processor\cite{mochida2018analog}}&180 nm& 2 M & analog & 0.33 & 15.8 & 20.7\\
      \cline{2-7}
      &40 nm & 4 M&analog&0.66 & 9.9 & 66.5\\
      \hline
      Fully Integrated Analog Chip\cite{liu2020integrated}& 130 nm & 4 K  & fixed-1/tenary&-&-&78.4\\
      \hline
      \multirow{5}{*}{Proposed}&\multirow{5}{*}{45 nm}&\multirow{5}{*}{256*512}&fixed-3/fixed-1&1524&1.15&1325.22\\
      \cline{4-7}
      &&&fixed-2/fixed-2&1092.2&0.77&1418.44\\
      \cline{4-7}
      &&&fixed-3/fixed-2&1092.2&1.16&941.55\\
      \cline{4-7}
      &&&fixed-4/fixed-4&546.1&1.47&371.49\\
      \cline{4-7}
      &&&fixed-8/fixed-8&121.4&2.00&60.68\\
      \hline
    \end{tabular}
    }
    \label{Table:core_comp}
\end{table*}

\subsection{Performance}
Table \ref{Table:core_performance} shows the dynamic performance comparison between the MBRAI \cite{zhang2020robust},  RPN\&BLM \cite{zhang2020regulator} and the proposed scheme. The M-RD4 recoder has a supply voltage of 0.6 V to further decreases the power consumption. The neuron circuit's supply voltage is 1 V to ensure the robustness of our proposed scheme. The computing speed, SFDR, SNDR, Effective Number of Bits (ENOB) of our proposed scheme are 1.85 M/s, 63.41 dB, 46.48 dB, and 7.42 bit, which are slightly better than the others. Table \ref{Table:energy_cost} gives the energy cost comparison of MBRAI, MBHS-mCNN \cite{yao2020fully}, RPN\&BLM, and our proposed scheme. MBRAI consumes 0.22 mW on amplifiers for stable read voltage, which means that amplifiers consume more than 90\% power, resulting in total power consumption is 199.68 mW. The ADCs consume more than 85\% energy in MBHS-mCNN, while the power consumption for $128\times256$ core is 7.44 mW.  RPN\&BLM uses regulators, with 1.11 uW power consumption, to keep the read voltage stable, and the total power consumption is 3.61 mW. 
In contrast, the power consumption of our proposed core is only 2.00 mW. Compared with MBRAI, MBHS-mCNN, and RPN\&BLM, the power consumption of our proposed scheme is reduced by 98.9\%, 73.1\% and 44.6\%, respectively.

The core level comparison between our proposed scheme and the other CIM core schemes is shown in Table \ref{Table:core_comp}. The simulation results show that our proposed design achieves energy efficiency as high as 60.68 TOPS/s/W in 8-bit input 8-bit weight pattern, 371.49 TOPS/s/W in 4-bit input 4-bit weight pattern, 941.55 TOPS/s/W in 3-bit input 2-bit weight pattern, 1418.44 TOPS/s/W in 2-bit input 2-bit weight pattern, and 1325.22 TOPS/s/W in 3-bit input 1-bit weight pattern. Compared with the other schemes, our proposed scheme achieves much higher efficiency. In the 8-bit input 8-bit weight pattern, our proposed scheme achieves an efficiency which is 99.47 $\times$, 5.44 $\times$, and 1.80 $\times$ more efficient than MBRAI, MBHS-mCNN, and RPN\&BLM schemes. Compared with other CIM schemes, our proposed CIM core achieves better energy efficiency.

\begin{table}[htpb]
\centering
\caption{Accuracy estimate of different RRAM-based schemes}
\begin{tabular}{|c|c|c|}
    \hline
        Network & Structure & Top-1 Error Rate  \\
        \hline
        \multirow{5}{*}{LeNet on MNIST}& Software Based & 0.90 \%\\
        \cline{2-3}
        &MBRAI&0.97 \%\\
        \cline{2-3} 
        &MBHS-mCNN&2.44 \%\\
        \cline{2-3}
        &RPN\&BLM&0.90 \%\\
        \cline{2-3}
        &Proposed&\textbf{ 0.91 \%}\\
        \hline
        \multirow{4}{*}{AlexNet on ILSVRC12}& Software Based & 42.70 \%\\
        \cline{2-3}
        &MBRAI&44.16 \%\\
        \cline{2-3}
        &RPN\&BLM&43.60 \%\\
        \cline{2-3}
        &Proposed&\textbf{43.10 \%}\\
        \hline
         \multirow{2}{*}{ResNet34 on ILSVRC12}& Software Based & 26.70 \%\\
         \cline{2-3}
        &Proposed&\textbf{27.80 \%}\\
        \hline
          \multirow{2}{*}{VGG16 on ILSVRC12}& Software Based & 28.40 \%\\
         \cline{2-3}
        &Proposed&\textbf{29.30 \%}\\
        \hline
    \end{tabular}
 \label{Table:network_accuracy}
\end{table}

\begin{table*}[!htpb]
    \centering
    \caption{Energy estimate of different RRAM-based schemes}
    \resizebox{\textwidth}{28mm}{
    \begin{tabular}{|c|c|c|c|c|c|c|c|c|}
        \hline
       \multirow{2}{*}{Network} &Number of& \multirow{2}{*}{Structure} &Ratio of&\multirow{2}{*}{System Frequency}& \multirow{2}{*}{Data Bit} & \multirow{2}{*}{Crossbar Size}& Energy&Saving \\
       &Operations&&1$\times$1&&&&(uJ/img)&\% \\
       \hline
       \multirow{2}{*}{LeNet on  } & \multirow{4}{*}{0.42 M }& MBRAI\cite{zhang2020robust}&\multirow{3}{*}{0.147}& 25 MHz & 8 & 256*256 & 0.71 & \textbf{98.9 \%} \\
       \cline{3-3}
       \cline{5-9}
        & & MBHS-mCNN\cite{yao2020fully} &&  25 MHz & 8 & 128*256 & 0.039 & \textbf{81.6 \%} \\
       \cline{3-3}
       \cline{5-9}
        \multirow{2}{*}{MNIST}& & RPN \& BLM\cite{zhang2020regulator}& & 16.7 MHz & 8 & 256*256 & 0.013 & \textbf{44.6 \%} \\
       \cline{3-9}
       & & Proposed & 0.022&16.7 MHz & 8 & 256*512 & \textbf{7.19E-3} &  - \\
       \hline
       \multirow{2}{*}{AlexNet on} & \multirow{4}{*}{720 M} & MBRAI\cite{zhang2020robust}&\multirow{3}{*}{0.143} & 25 MHz &8& 256*256& 1.23E+03 & \textbf{98.9 \%} \\
       \cline{3-3}
       \cline{5-9}
       & &  MBHS-mCNN\cite{yao2020fully} & &25 MHz & 8 & 128*256 &68.56 & \textbf{81.5 \% }\\
       \cline{3-3}
       \cline{5-9}
        \multirow{2}{*}{ILSVRC2012}& & RPN \& BLM\cite{zhang2020regulator} && 16.7 MHz & 8 & 256*256 & 22.46 & \textbf{43.6 \%} \\
       \cline{3-9}
       & & Proposed & 0.029&16.7 MHz & 8 & 256*512 & \textbf{12.66} & - \\
       \hline
       \multirow{2}{*}{ResNet34 on } & \multirow{4}{*}{4 G} & MBRAI\cite{zhang2020robust} &\multirow{3}{*}{0.125}& 25 MHz & 8 & 256*256 & 6.92E+03 & \textbf{98.9 \%} \\
       \cline{3-3}
       \cline{5-9}
       & & MBHS-mCNN\cite{yao2020fully} && 25 MHz & 8 & 128*256 & 390.07 & \textbf{81.1 \%} \\
      \cline{3-3}
       \cline{5-9}
       \multirow{2}{*}{ILSVRC2012}  & & RPN \& BLM\cite{zhang2020regulator} && 16.7 MHz & 8 & 256*256 & 141.95 & \textbf{48.1 \%} \\
       \cline{3-9}
       & & Proposed &0.037& 16.7 MHz & 8 & 256*512 & \textbf{73.73} & - \\
       \hline
       \multirow{2}{*}{VGG16 on }&\multirow{4}{*}{16 G}& MBRAI\cite{zhang2020robust} &\multirow{3}{*}{0.129}& 25 MHz &8& 256*256& 2.77E+04 & \textbf{98.9 \%} \\
       \cline{3-3}
       \cline{5-9}
       & & MBHS-mCNN\cite{yao2020fully} && 25 MHz & 8 & 128*256 &1.56E+03& \textbf{81.9 \%} \\
       \cline{3-3}
       \cline{5-9}
        \multirow{2}{*}{ILSVRC2012}&  & RPN \& BLM\cite{zhang2020regulator} && 16.7 MHz & 8 & 256*256 & 567.8 & \textbf{50.3 \%} \\
       \cline{3-9}
       & & Proposed &0.022& 16.7 MHz & 8 & 256*512 & \textbf{282.35}& - \\
       \hline
       \end{tabular}
  }
    \label{Table: network_energy}
\end{table*}

\subsection{Network-Level Estimation}
To estimate the accuracy and energy estimate of our proposed scheme, the model of LeNet \cite{lecun1998gradient}  on the dataset MNIST and the models of AlexNet \cite{krizhevsky2017imagenet}, ResNet34 \cite{he2016deep} and VGG16 \cite{simonyan2014very} on ILSVRC2012 are evaluated with the mapping method mentioned in section III.D. The estimated accuracy is shown in Table \ref{Table:network_accuracy}. Our proposed scheme achieves an accuracy better than MBHS-mCNN in LeNet, and roughly equivalent to MBRAI and RPN\&BLM in LeNet and AlexNet. 
The energy estimation between the proposed scheme and other RRAM based schemes is shown in Table \ref{Table: network_energy}. The model of LeNet on the dataset MNIST is used to test the performance of the schemes in small-scale networks. The models of AlexNet, ResNet34 and VGG16 on ILSVRC2012 are used to evaluate the performance in large-scale networks. Our proposed scheme reduces the ratio of $1\times1$ by 78.5\% on LeNet, 80.2\% on AlexNet, 70.4\% on ResNet34 and 82.9\% on VGG16. Therefore, the power consumption is greatly reduced. The inference energy per image is reduced by 98.9\% compared with MBRAI, more than 81.5\% compared with MBHS-mCNN, and more than 43.6\% compared with RPN\&BLM on different nerworks.  Therefore, the inference energy is significantly reduced in our proposed scheme  by abandoning the amplifiers and adopting M-RD4 and M-CSD codes.


\begin{figure}[htpb]
    \centering
    \includegraphics[width=0.5\textwidth]{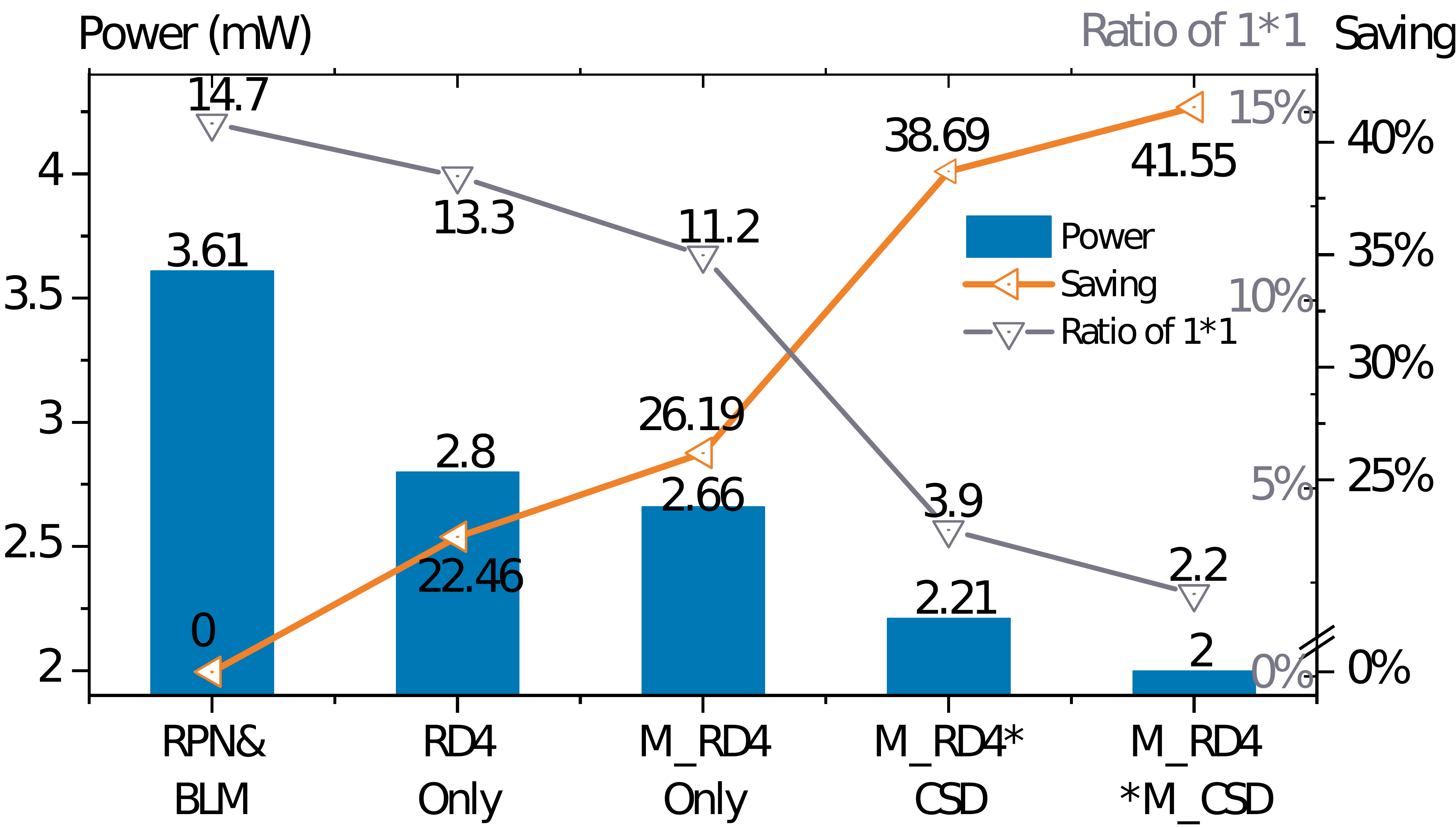}
    \caption{The energy cost comparison between different code combination.}
    \label{fig:code_comp}
\end{figure}

As shown in Fig. \ref{fig:code_comp},  the energy cost and the ratio of 1$\times$1 of different codes are simulated to verify the superiority of out proposed M-RD4 and M-CSD. The core size is set to be 256*512, and the ratio of 1$\times$1 is obtained on LeNet with the dataset MNIST. Take  RPL\&BLM as the standard, the ratio of 1$\times$1 with binary input and binary weight is 14.7\% and the power consumption is 3.61 mW. The ratio of 1$\times$ decreases to 13.3\% by using radix-4 input. What's more, the ratio of 1$\times$1 is further decreased to 11.2\% by using our M-RD4 in the input, and the power consumption is decreased by 26.46\%, respectively. Applying the M-RD4 input and CSD weight, the ratio of 1$\times$1 decreases to 3.9\%, and the power consumption decreases to 2.21 mW. Our proposed scheme with M-RD4 input and M-CSD weight further decreases the ratio of 1$\times$1 to 2.2\% and the power consumption to 2.00 mW. Therefore, for a 256*512 core, our proposed scheme saves 41.55\% of power consumption compared with RPN\& BLM.

\section{Conclusion}
In this paper, a low power in-memory multiplication and accumulation array with modified radix-4 input and canonical-signed-digit weights has been proposed. Modified radix-4 booth code is used to reduce the number of `1's in the input data, and differential memory pairs with modified canonical-signed-digit are used to reduce the `1's in weight. The proposed two coding schemes efficiently reduce the ratio of $1\times1$ by 85.0\% on LeNet, 79.7\% on AlexNet, 70.4\% on ResNet34 and 82.9\% on VGG16. The simulation results has shown that our proposed CIM core achieves 2.00 mW on power consumption with 256*512 in 8-bit input and 8-bit weight pattern. The computing-power rate at the fixed-point 8-bit is 60.68 TOPS/s/W, which is 99.47$\times$, 5.44$\times$, and 1.80$\times$ than that of MBRAI, MBHS-mCNN and RPN\&BLM schemes, respectively. The core is very robust with an ENOB of 7.42-bit whose SFDR and SNDR achieve 63.41 dB and 46.48 dB. The network-level estimation has shown that the proposed core achieves 0.91\% top-1 error rate with 7.59E-3 uJ/img on LeNet, 43.60\% top-1 error rate with 13.36 uJ/img on AlexNet, 27.80\% top-1 error rate with 77.79 uJ/img on ResNet34, and 29.30\% top-1 error rate with 297.88 uJ/img on VGG16, respectively. The core achieves very low inference energy cost and high accuracy, which are much better than other schemes. The linearity and PVT simulation has been done to verify the robustness of the circuit. The energy efficiency comparison has shown that the proposed scheme achieves much lower power consumption than others.



\bibliographystyle{IEEEtran}
\bibliography{bib.bib}

\begin{IEEEbiography}[{\includegraphics[width=1in,height=1.25in,clip,keepaspectratio]{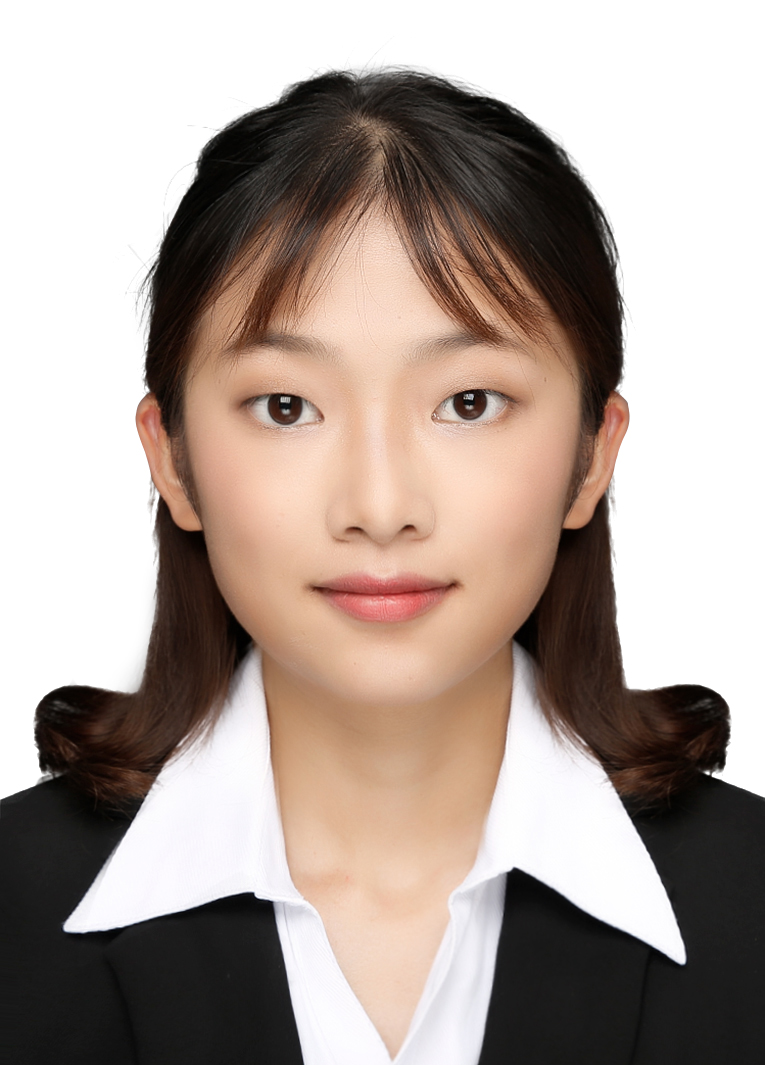}}]
{Rui Xiao}(Student Member, IEEE) received the Bechalor degree from the College of Information Science Electronic Engineering, Zhejiang University in 2019. Currently she is pursuing the Ph.D degree in the School of Information Science and Electronic Engineering, Zhejiang University under the supervision of Prof. Huang. Her research interests include in-memory computing circuits and systems design using emerging resistive non-volatile memories, deep learning accelerators, and embedded system design.
\end{IEEEbiography}

\begin{IEEEbiography}[{\includegraphics[width=1in,height=1.25in,clip,keepaspectratio]{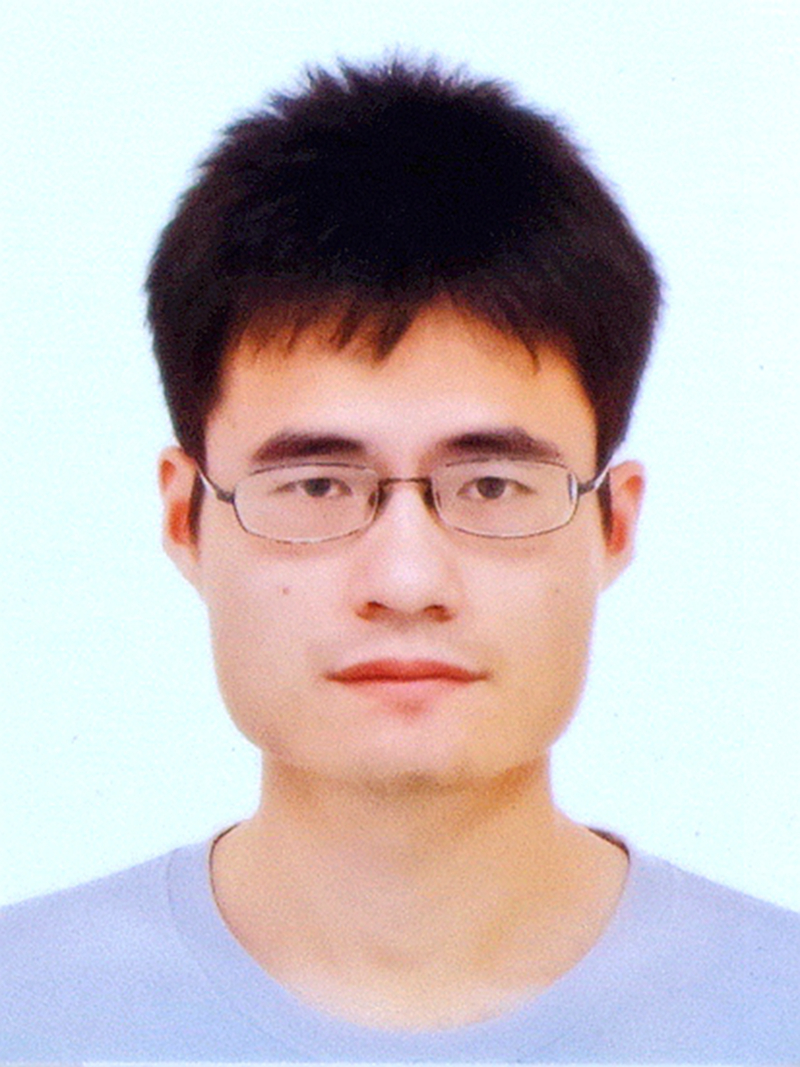}}]
{Kejie Huang}(Senior Member, IEEE) received the Ph.D. degree from the Department of Electrical Engineering, National University of Singapore (NUS), Singapore, in 2014. He has been a Principal Investigator with the College of Information Science Electronic Engineering, Zhejiang University (ZJU), since 2016. Prior to joining ZJU, he has spent five years at the IC design industry, including Samsung and Xilinx, two years in the Data Storage Institute, Agency for Science Technology and Research (A*STAR), and another three years in the Singapore University of Technology and Design (SUTD), Singapore. He has authored or coauthored 40 scientific articles in international peer-reviewed journals and conference proceedings. He holds four granted international patents, and another eight pending ones. His research interests include low power circuits and systems design using emerging non-volatile memories, architecture and circuit optimization for reconfigurable computing systems and neuromorphic systems, machine learning, and deep learning chip design. He currently serves as the Associate Editor of the IEEE TRANSACTIONS ON CIRCUITS AND SYSTEMS-PART II: EXPRESS BRIEFS.
\end{IEEEbiography}

\begin{IEEEbiography}[{\includegraphics[width=1in,height=1.25in,clip,keepaspectratio]{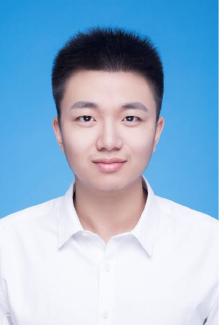}}]{Yewei Zhang}
(Student Member, IEEE) recieved the bachelor’s degree from College of Information Science \& Electronic Engineering, Zhe Jiang University in 2018. He is currently studying for a master's degree at College of Information Science \& Electronic Engineering, Zhe Jiang University. He is interested in in-memory computing and non-volatile memories.
\end{IEEEbiography}

\begin{IEEEbiography}[{\includegraphics[width=1in,height=1.25in,clip,keepaspectratio]{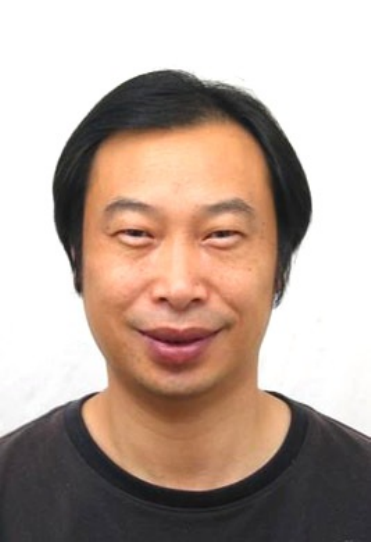}}]{Haibin Shen}
is currently a Professor with Zhejiang
University, a member of the second level of 151 talents
project of Zhejiang Province, and a member
of the Key Team of Zhejiang Science and Technology
Innovation. His research interests include learning
algorithm, processor architecture, and modeling.
His research achievement has been used by many
authority organizations. He has published more than
100 papers on academic journals, and he has been
granted more than 30 patents of invention. He was a
recipient of the First Prize of Electronic Information
Science and Technology Award from the Chinese Institute of Electronics, and
has won a second prize at the provincial level.
\end{IEEEbiography}

\end{document}